\documentclass[preprint,nofootinbib,amsmath,amssymb,aps,prd]{revtex4}
\usepackage{braket}             
\usepackage{graphicx}           
\usepackage{bm}             
\usepackage[usenames, dvipsnames]{color}
\usepackage{hyperref}           
\usepackage{mathtools}
\usepackage[normalem]{ulem} 
\usepackage[caption=false]{subfig}
\usepackage{float}
\usepackage{graphicx}
\usepackage{soul}              
\DeclareMathOperator{\tr}{tr}           

\renewcommand{\Im}{\mathop{\rm Im}} 

\usepackage{colonequals} 

\usepackage[usenames, dvipsnames]{xcolor}

\definecolor{capri}{rgb}{0.0, 0.75, 1.0}

\begin{document}

\title{Harvesting correlations from vacuum quantum fields in the presence of a reflecting boundary}
\author{Zhihong Liu$^{1}$~\footnote{zhihongliu@hunnu.edu.cn}, Jialin Zhang$^{1,2}$~\footnote{Corresponding author at jialinzhang@hunnu.edu.cn}and Hongwei Yu$^{1,2}$~\footnote{Corresponding author at hwyu@hunnu.edu.cn}}
\affiliation{$^1$ Department of Physics and Synergetic Innovation Center for Quantum Effects and Applications, Hunan Normal University, 36 Lushan Rd., Changsha, Hunan 410081, China\\
$^2$ Institute of Interdisciplinary Studies, Hunan Normal University, 36 Lushan Rd., Changsha, Hunan 410081, China}
\date{\today}
\begin{abstract}
We explore correlations harvesting  by two static detectors locally interacting with vacuum massless scalar fields in the presence of an infinite perfectly reflecting boundary. We study the phenomena of mutual information harvesting and entanglement  harvesting for two detector-boundary alignments, i.e., parallel-to-boundary  and orthogonal-to-boundary alignments.   Our results show that the presence of the boundary generally inhibits  mutual information harvesting relative to that in flat spacetime without any boundaries. In contrast,  the boundary may play a doubled-edged role in entanglement harvesting, i.e., inhibiting entanglement harvesting in the near zone of the boundary while assisting it in the far zone of the boundary. Moreover,  there exists an optimal detector energy gap difference between two nonidentical detectors  that makes such detectors  advantageous  in correlations harvesting as long as the interdetector separation is large enough. The value of the optimal detector energy gap difference depends on both the interdetector separation  and the detector-to-boundary distance.  A comparison of the correlations harvesting in two different alignments shows that although correlations harvesting share qualitatively the same properties, they also display quantitative differences in that the detectors in orthogonal-to-boundary alignment always harvest comparatively more mutual information than the parallel-to-boundary ones, while they  harvest comparatively more entanglement  only near the boundary.

\end{abstract}

\maketitle


\section{Introduction}
It has been recognized  for a  long time  that the vacuum state of free quantum fields possesses  nonlocal properties and contains   correlations between timelike and spacelike separated regions~\cite{Summers:1985, Summers:1987,Pozas-Kerstjens:2015,Ng:2018}. Such correlations, which may be either quantum entanglement, quantum mutual information or quantum discord~\cite{M.Naeem:2022}, can be extracted by a pair of initially uncorrelated Unruh-DeWitt (UDW) particle detectors interacting with the vacuum quantum fields.
This phenomenon has been known as correlation  harvesting, and such  harvesting process is  referred  to  as the correlation harvesting protocol~\cite{Pozas-Kerstjens:2015}.

The correlation harvesting   in terms of quantum entanglement has been extensively studied in different circumstances in  recent years and has been demonstrated to be  sensitive to the  properties of spacetime including  dimensionality~\cite{Pozas-Kerstjens:2015}, topology~\cite{EDU:2016-1}, curvature~\cite{Steeg:2009,Nambu:2013,Kukita:2017,Ng:2018,Ng:2018-2,Zhjl:2018,Zhjl:2019,Robbins:2020jca,Gallock-Yoshimura:2021} and the presence of boundaries~\cite{CW:2020,Zhjl:2021,CW:2019}, the  intrinsic motion and energy gap of detectors~\cite{Salton-Man:2015,Zhjl:2020,Zhjl:2021,Zhjl:2022,R.B.Mann:2022,Maeso:2022,Zhjl:2022.08}, and superpositions of temporal order for detectors~\cite{Henderson:2020}.

Recently, the study of entanglement  harvesting has been extended to nonidentical detectors with different energy gaps  in  flat spacetime~\cite{Zhjl:2022.4}, in contrast to previous studies where detectors are usually assumed  to be identical. It was argued that the presence of energy gap difference generally enlarges the rangefinding of entanglement harvesting  (the harvesting-achievable range for the interdetector separation), and  two nonidentical detectors can  harvest more entanglement than the identical ones if the interdetector separation is not too small with respect to the interaction duration time. However,  the validity of such results in other  circumstances such as in  curved spacetime or in the presence of boundaries  merits further exploration.  In fact, it has been recognized that the presence of reflecting boundaries in flat spacetime 
modifies the fluctuations of quantum fields, resulting in some intriguing quantum effects
, such as the Casimir effect~\cite{Casimir:1948},  the light-cone fluctuations~\cite{Yu:1999}, the geometric phase~\cite{Zhjl:2016} and the modification for the radiative properties of accelerated atoms~\cite{Yu:2005,Yu:2006,Rizzuto:2009}.  Since the dynamic evolution in time of the detectors system 
 is strongly dependent on the fluctuations of quantum fields,   modifications of the fields fluctuations caused by the presence of a boundary have been shown to  play a significant role in controlling the entanglement creation in the detectors system~\cite{Zhang:2007,Zhang:2007.1,Cheng:2018}.  And the phenomena  of  entanglement harvesting by two UDW detectors locally interacting with the fluctuating  quantum fields have also been examined in the  presence  of a reflecting boundary  but in a simple scenario: two identical detectors with not too large energy gap in parallel-to-boundary alignment~\cite{Zhjl:2021}. It was found, through the numerical evaluation, that the reflecting  boundary plays a double-edged role in entanglement harvesting, i.e., degrading  the harvested entanglement amount in general while enlarging the entanglement harvesting-achievable interdetector separation range. However, what happens to
 the entanglement harvesting phenomenon in the presence of  a  boundary  when  detectors  are nonidentical remains to be investigated.

On the other hand, the correlation harvesting   in terms of mutual information, a useful measure on information which  quantifies the total amount of classical and quantum correlations including entanglement, has also been recently studied.
It has been found that the mutual information harvesting  also depends upon the intrinsic  properties of spacetime, the detectors' energy gap and noninertial  motion~\cite{Pozas-Kerstjens:2015,Simidzija:2018,Kendra:2022,M.Naeem:2022}. Remarkably, unlike the entanglement harvesting  that has a finite harvesting-achievable separation range and cannot occur near a black hole,  mutual information harvesting can occur everywhere with an arbitrarily large interdetector separation~\cite{Pozas-Kerstjens:2015} and even does not vanish near the event horizon of a black hole at an extremely high local Hawking temperature~\cite{Kendra:2022}. However, in comparison to the entanglement harvesting which is relatively well understood, much remains to be done on understanding the phenomenon of mutual information harvesting,  for example, the mutual information harvesting by two identical/nonidentical detectors near a boundary.

In this paper, we will perform a general study of the correlation harvesting phenomenon for two detectors placed near a reflecting boundary, focusing upon entanglement harvesting and mutual information harvesting phenomenon  by nonidentical detectors. Our particular interest lies in the influence of energy gaps on mutual information harvesting and the role played by a perfectly reflecting plane boundary, 
 including correlation harvesting  by nonidentical detectors with different energy gaps in the scenarios of parallel-to-boundary and orthogonal-to-boundary alignments.

 The paper is organized as follows.   We begin in section II by briefly reviewing  the UDW detector model, the correlation harvesting protocol, and introduce the conventional  measures  for entanglement and mutual information. In section III, we respectively explore the phenomena of entanglement harvesting and mutual information  harvesting for two static detectors aligned parallel to the boundary, where the influence of the boundary and detectors' energy gaps on correlation harvesting will be studied in detail, and make a qualitative comparison between the phenomena of mutual information harvesting and entanglement harvesting. In section IV, the correlation harvesting phenomenon for detectors orthogonally  aligned to the boundary  will be studied, and some comparisons of the results between the parallel-to-boundary and the orthogonal-to-boundary alignments are made.  Finally, we end  with conclusions in section V. Throughout the paper,  the natural units $\hbar=c=k_B=1$ are adopted for convenience.

\section{The basic formalism}

We consider two  detectors $A$ and $B$  locally interacting with a massless quantum scalar field $\phi[x_D(\tau)]$ {($D\in\{A,B\}$)} along their worldlines. The spacetime trajectory of the detector, $x_{D}(\tau)$,  is parametrized by its proper time $\tau$.  The detector-field coupling is given by the following interaction Hamiltonian
\begin{equation}\label{Int1}
 H_{D}(\tau)=\lambda \chi(\tau)\left[e^{i \Omega_{D}\tau} \sigma^{+}+e^{-i \Omega_{D}\tau} \sigma^{-}\right] \phi\left[x_{D}(\tau)\right],~~ D\in\{A,B\}\;,
\end{equation}
where $\lambda$ is the coupling strength, and $\chi(\tau)=\exp[-{\tau^{2}}/(2\sigma^{2})]$ is the Gaussian switching function with parameter $\sigma$ controlling the duration time of the interaction. In principle, all relevant physical quantities  can be rescaled with the duration time parameter $\sigma$ to be unitless. Here, we use the standard UDW detector model to describe a two-level
system with an energy gap $\Omega_{D}$  between its ground state $|0\rangle_{D}$ and  excited state $|1\rangle_{D}$. The operators $\sigma^{+}=|1\rangle_{D}\langle0|_{D}$ and $\sigma^{-}=|0\rangle_{D}\langle1|_{D}$ are just the SU(2) ladder operators.

Suppose the two detectors are prepared in their ground state and the field is in the Minkowski vacuum state $|0\rangle_{M}$, then the initial state of the two detectors and field system  is given by $|\Psi\rangle_{i}=|0\rangle_{A}|0\rangle_{B}|0\rangle_{M}$.  The time-evolution of the  quantum system can be obtained by using the Hamiltonian~(\ref{Int1}), 
\begin{equation}\label{psi-f}
|\Psi\rangle_{f}:={\cal{T}} \exp\Big[-i\int{dt}\Big(\frac{d\tau_A}{dt}H_A(\tau_A)+
\frac{d\tau_B}{dt}{H_B}(\tau_B)\Big)\Big]|\Psi\rangle_{i}\;,
\end{equation}
where ${\cal{T}}$ is the time ordering operator and $t$ is the coordinate time with respect to which the vacuum state of the field is defined.
By tracing out the field degrees of freedom in Eq.~(\ref{psi-f}),  the  density matrix for the final state of the two detectors turns, to leading order in the interaction strength and in the
basis $\{|0\rangle_{A}|0\rangle_{B},|0\rangle_{A}|1\rangle_{B},|1\rangle_{A}|0\rangle_{B},
|1\rangle_{A}|1\rangle_{B}\}$, out to be~\cite{Pozas-Kerstjens:2015,Zhjl:2019}
\begin{align}\label{rhoAB}
\rho_{AB}:&=\tr_{\phi}\big(U \ket{\Psi_{f}}\bra{\Psi_{f}} U^{\dag}\big)\nonumber\\
 &=\begin{pmatrix}
 1-P_A-P_B & 0 & 0 & X \\
 0 & P_B & C & 0 \\
 0 & C^* & P_A & 0 \\
 X^* & 0 & 0 & 0 \\
 \end{pmatrix}+{\mathcal{O}}(\lambda^4)\;,
\end{align}
where the detector transition probability reads
\begin{equation}\label{PAPB}
 P_D:=\lambda^{2}\iint d\tau d\tau' \chi(\tau) \chi(\tau') e^{-i \Omega_{D}(\tau-\tau')}
 W\left(x_D(t), x_D(t')\right)\quad\quad D\in\{A, B\}\;,
\end{equation}
and the quantities $C$ and $X$  which characterize correlations of the two detectors, are given by
\begin{align}\label{ccdef}
C &:=\lambda^{2} \iint d \tau d \tau^{\prime} \chi(\tau) \chi(\tau')
 e^{-i (\Omega_{A}\tau-\Omega_{B}\tau')} W\left(x_{A}(t), x_{B}(t')\right)\;,
\end{align}
\begin{align}\label{xxdef}
X:=&-\lambda^{2} \iint d\tau d \tau' \chi(\tau)\chi(\tau') e^{-i(\Omega_{A}\tau+\Omega_{B}\tau')}
 \Big[\theta(t'-t)W\left(x_A(t),x_B(t')\right)\nonumber\\&+\theta(t-t')W\left(x_B(t'),x_A(t)\right)\Big]\;.
\end{align}
Here $W(x,x'):=\langle0|_{M}\phi(x)\phi(x')|0\rangle_{M}$ is the Wightman function of the quantum field in the Minkowski vacuum state,
and $\theta(t)$ is the Heaviside theta function. In particular, if detectors are at rest, one has $t=\tau$, i.e., the coordinate time of the detector is equal to its proper time.
The amount of quantum entanglement can be measured  by concurrence~\cite{WW}, which, with the density matrix~(\ref{rhoAB}), is given by~\cite{EDU:2016-1,Zhjl:2018,Zhjl:2019}
\begin{equation}\label{condf}
\mathcal{C}(\rho_{A B})=2 \max \Big[0,|X|-\sqrt{P_{A}
P_{B}}\Big]+\mathcal{O}(\lambda^{4})\;.
\end{equation}
Obviously, the concurrence is a competition between the correlation term $X$ and geometric mean of the detectors' transition probabilities $P_{A}$ and $P_{B}$.
The total amount of correlations is characterized by mutual information, which is defined as~\cite{Nielsen:2000}
\begin{equation}\label{IIdf0}
\mathcal{I}(\rho_{AB})= S(\rho_{A})+S(\rho_{B})-S(\rho_{AB})\;,
\end{equation}
where $\rho_{A}=\operatorname{tr}_{B}(\rho_{AB})$ and $\rho_{B}=\operatorname{tr}_{A}(\rho_{AB})$ denote the  partial
traces of detectors' state $\rho_{AB}$,  and $S(\rho)=-\operatorname{tr}(\rho\ln\rho)$ is the von Neumann entropy.  With the above definition~(\ref{IIdf0}), the mutual information for the  final detector state~(\ref{rhoAB}) takes the following form~\cite{Pozas-Kerstjens:2015}
\begin{align}\label{IIdf}
\mathcal{I}(\rho_{AB})=&\mathcal{L}_+\ln(\mathcal{L}_{+})+\mathcal{L}_-\ln(\mathcal{L}_{-})-P_{A}\ln(P_{A})-P_{B}\ln (P_{B})+\mathcal{O}(\lambda^4)
\end{align}
with
\begin{align}
\mathcal{L}_\pm&=\frac{1}{2}\Big(P_{A}+P_{B}\pm\sqrt{ (P_{A}-P_{B})^2 + 4 |C|^2 }\Big)\;.
\end{align}
In  contrast to  the concurrence~(\ref{condf}), the  mutual information is  determined by  transition probabilities $P_{A}$, $P_{B}$ and the correlation term $C$ rather than the correlation term $X$. It  can be  verified that mutual information is an increasing function of  parameter $|C|$, and monotonically decreases as the transition probabilities increase.  According to Eq.~(\ref{IIdf}), we can further obtain ${\cal {I}}(\rho_{AB})=0$ if $C=0$.  Especially, when
a transition probability is zero, the  mutual information must vanish from  the positivity condition of the density matrix, (i.e., $P_{A}P_{B}\geq |C|^2$).

In what follows, we will explore, with the  harvesting protocol,  the correlations harvested by two static detectors near a perfectly reflecting boundary,  figuring out the role played by the presence of the  boundary  in both entanglement harvesting and mutual information harvesting. It is supposed that a  perfectly reflecting plane boundary is located at $z=0$, and  two nonidentical UDW detectors with different energy gaps are  aligned  in two different ways:  parallel-to-boundary and orthogonal-to-boundary.
\section{Correlation Harvesting for the detectors aligned parallel to the boundary}
\begin{figure}[!htbp]
  \centering
  {\includegraphics[width=0.40\linewidth]{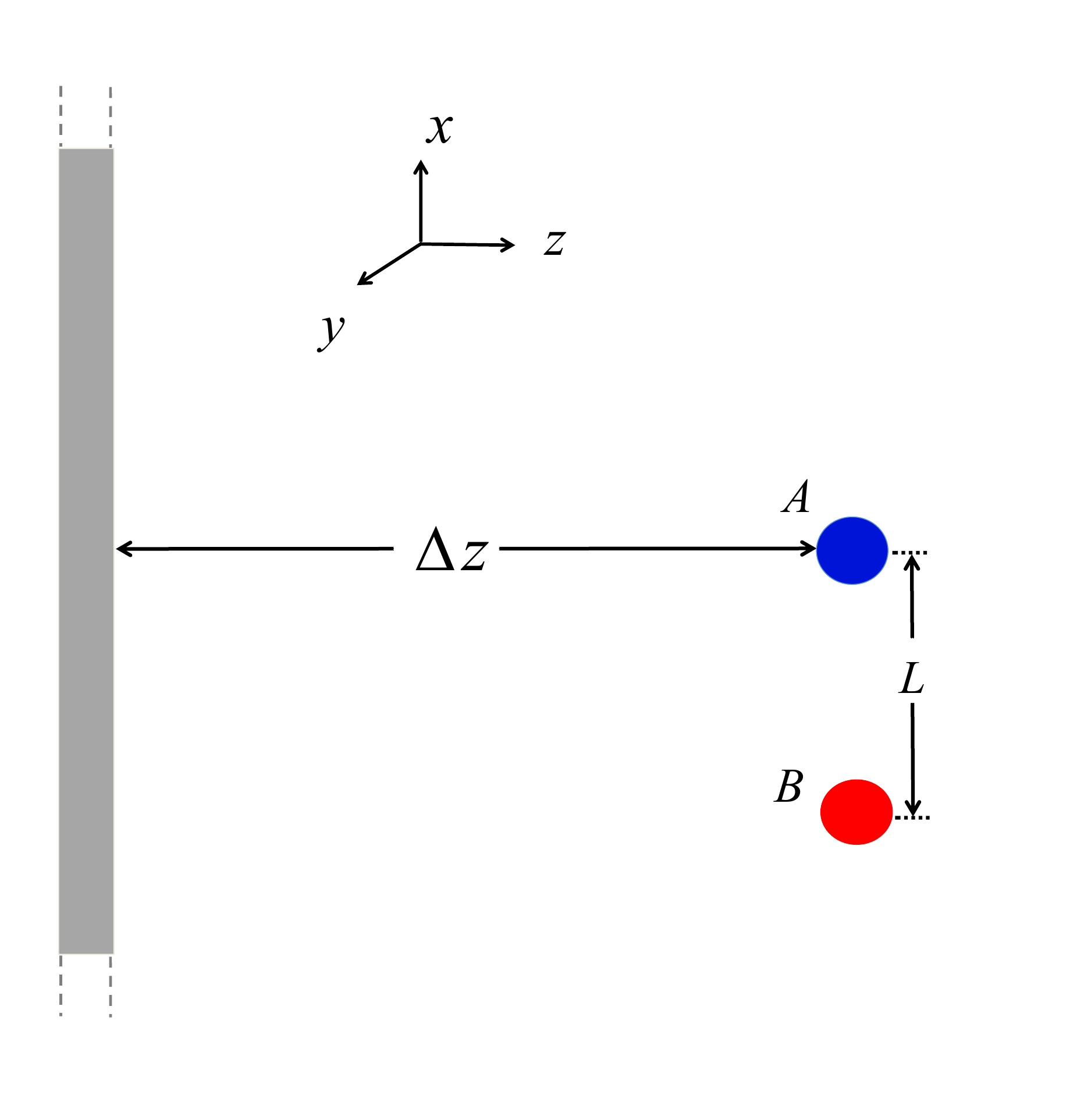}}\qquad\qquad\qquad
 \caption{ Two static detectors separated from each other by a distance $L$ are aligned parallel to the boundary at a distance $\Delta {z}$ away from the boundary.}
  \label{parmodel}
\end{figure}

In section, we  consider  that  two static detectors separated by a distance $L$ are aligned parallel to the boundary with a distance $\Delta{z}$ from the boundary (see Fig.~(\ref{parmodel})).  The spacetime trajectories of the detectors  can then be written as
\begin{align}\label{Static-trj}
x_A:=\{t_A\;,x=0\;,y=0\;,z=\Delta{z}\}\;,~~ x_B:=\{t_B\;,x=L\;,y=0\;,z=\Delta{z}\}\;.
\end{align}

Let us now calculate the transition probabilities $P_D$ given by Eq.~(\ref{PAPB}) and the correlation terms $C$ and $X$  respectively given by Eq.~(\ref{ccdef}) and Eq.~(\ref{xxdef}).  To do this, we need
the Wightman function for vacuum massless scalar fields, which is, according to the method of images, given by~\cite{Birrell:1984}
\begin{align}\label{wigh-2}
W\left(x_D, x'_D\right)=&-\frac{1}{4 \pi^{2}}\Big[\frac{1}{(t-t'-i
\epsilon)^{2}-(x-x')^{2}-(y-y')^{2}
-(z-z')^{2}}\nonumber\\
&-\frac{1}{(t-t'-i \epsilon)^{2}-(x-x')^{2}-(y-y')^{2}
-(z+z')^{2}}\Big]\;.
\end{align}
Substituting  trajectory~(\ref{Static-trj}) and  Eq.~(\ref{wigh-2}) into
Eq.~(\ref{PAPB}), one may straightforwardly  obtain the transition probabilities~\cite{Zhjl:2021}
\begin{align}\label{PD0}
P_{D} =&\frac{\lambda^{2}}{4 \pi}\left[e^{-\Omega_{D}^{2} \sigma^{2}}-\sqrt{\pi}
\Omega_{D} \sigma {\rm{Erfc}}(\Omega_{D}
\sigma)\right]-\frac{\lambda^{2}\sigma{e}^{-\Delta{z}^2/\sigma^2}}{8\sqrt{\pi}\Delta{z}}\nonumber\\
&\times\Big\{\Im\Big[e^{2i\Omega_{D}\Delta{z}}{\rm{Erf}}\Big(i\frac{\Delta{z}}{\sigma}+\Omega_{D}\sigma\Big)\Big]
-\sin(2\Omega_{D}\Delta{z})\Big\},\qquad D\in\{A,B\}\;,
\end{align}
where ${\rm{Erf}}(x)$ is the error function and $\operatorname{Erfc}(x):=1-{\rm{Erf}}(x)$.
Similarly, the  correlation terms $C$ and $X$ in this case can also be worked out
\begin{equation}\label{Expression-C}
    C=\frac{\lambda^{2}}{4\sqrt{\pi}}
    e^{-\frac{\Delta\Omega^{2}\sigma^{2}}{4}}\Big[f(L)-f(\sqrt{L^2+4\Delta{z}^2})\Big]\;,
\end{equation}
\begin{equation}\label{Expression-X}
    X=-\frac{\lambda^{2}}{4\sqrt{\pi}}
    e^{-\frac{(2\Omega_A+\Delta\Omega)^{2}\sigma^{2}}{4}}\Big[g(L)-g(\sqrt{L^2+4\Delta{z}^2})\Big]\;,
\end{equation}
where the auxiliary functions, $f$ and $g$, are defined by
\begin{equation}\label{Expression-C0}
f(L):=\frac{{\sigma} e^{-L^2/(4\sigma^{2})}}{L}\Big\{\Im \Big[e^{i(2\Omega_A+\Delta\Omega){L}/2}{\rm{Erf}}
\Big(\frac{iL+2\Omega_A\sigma^2+\Delta\Omega\sigma^2}{2\sigma}\Big)\Big]
    -\sin\Big[\frac{(2\Omega_A+\Delta\Omega)L}{2}\Big]\Big\}\;,
\end{equation}
and
\begin{equation}\label{Expression-X0}
   g(L):=\frac{{\sigma}e^{-L^2/(4\sigma^{2})}}{L}\Big\{\Im\Big[e^{i\Delta\Omega{L}/2}
    {\rm{Erf}}\Big(\frac{iL+\Delta\Omega\sigma^2}{2\sigma}\Big)\Big]
    +i\cos\Big(\frac{\Delta\Omega{L}}{2}\Big)\Big\}\;.
\end{equation}
Without loss of generality, we here assume the energy gap of the detector $B$ is not less than that of detector $A$, i.e., $\Delta\Omega:=\Omega_B-\Omega_A \geq 0$ throughout the paper. From Eq.~(\ref{Expression-C}) and Eq.~(\ref{Expression-X}), it is easy to find  that as long as the energy gap difference is large with respect to duration time parameter ($\Delta\Omega\gg1/\sigma$), both the correlation terms $C$ and $X$ become vanishingly small, and so  mutual information and  entanglement can hardly  be harvested.

\subsection{Entanglement harvesting}
In general, it is very difficult to capture the characteristics of  the correlation harvesting phenomena  with the very complicated analytical results above. So, the numerical evaluations are usually called for investigating the influence of the boundary and the energy gaps on correlation harvesting.  However, in some special cases, one can still perform  analytical approximations to the afore-results.   For  two identical
detectors, i.e., $\Omega_{A}=\Omega_{B}=\Omega$, the approximate results for harvested entanglement can be
derived in two cases: small energy gap ($\Omega\sigma\ll1$) and large energy gap ($\Omega\sigma\gg1$).

\subsubsection{Small energy gap}
When the detectors system is very close the boundary (i.e., $\Delta{z}/\sigma\ll1$), the transition probabilities  can be approximated as
\begin{align}\label{PD11}
P_{A}=P_{B}\approx\frac{\lambda^{2}\Delta{z}^{2}}{2\pi\sigma^{2}}\Big(\frac{1}{3}-\frac{\sqrt{\pi}\Omega\sigma}{2}\Big),
\end{align}
and the  correlation term $|X|$ as
\begin{equation}\label{Expression-x1}
 |X|\approx\left\{\begin{aligned}
&\frac{\lambda^{2}\Delta{z}^{2}\sigma}{2L^{3}\sqrt{\pi}}\Big(1-\Omega^{2}\sigma^{2}\Big),&\frac{L}{\sigma}\ll1;\\
&\frac{2\lambda^{2}\Delta{z}^{2}\sigma^{2}}{L^{4}\pi}\Big(1-\Omega^{2}\sigma^{2}\Big),&\frac{L}{\sigma}\gg1\;.
 \end{aligned} \right.
\end{equation}
The concurrence, denoted here by $\mathcal{C}_{P}(\rho_{A B})$ 
can be  approximately expressed as
\begin{equation}\label{Expression-c1}
 \mathcal{C}_{P}(\rho_{A B})\approx\left\{\begin{aligned}
&\frac{\lambda^{2}\Delta{z}^{2}}{\sqrt{\pi}}\Big(\frac{\sigma}{L^{3}}+\frac{1}{4L\sigma}-
\frac{1}{3\sqrt{\pi}\sigma^{2}}+\frac{\Omega}{2\sigma}\Big)
,&\quad\frac{\Delta{z}}{\sigma}\ll\frac{L}{\sigma}\ll1\;;\\
&\max \Bigg[0, -\frac{\lambda^{2}\Delta{z}^{2}}{3\pi\sigma^{2}}+\frac{4\lambda^{2}\Delta{z}^{2}\sigma^{2}}{L^{4}\pi}\Bigg]=0,
&\quad\frac{\Delta{z}}{\sigma}\ll1,\frac{L}{\sigma}\gg1\;.
 \end{aligned} \right.
\end{equation}
 Eq.~(\ref{Expression-c1}) shows that the concurrence vanishes in the limit of $\Delta{z}\rightarrow0$, i.e, the two detectors are located at the boundary, as well as  when $L/\sigma\gg1$, i.e., the interdetector separation is very large. This means that entanglement harvesting cannot occur in these cases.   While for small interdetector separation  ($L/\sigma\ll1$), the harvested entanglement is an increasing function of the energy gap ($\Omega$)  and the detector-to-boundary distance ($\Delta{z}$). In the far zone to the boundary ($\Delta{z}/\sigma\gg1$), one can also approximately estimate the concurrence 
\begin{equation}\label{Expression-c2}
 \mathcal{C}_{P}(\rho_{A B})\approx\left\{\begin{aligned}
&\frac{\lambda^{2}}{2\sqrt{\pi}}\Big(\frac{\sigma}{L}-\frac{1}{\sqrt{\pi}}+\frac{\sigma^{2}}{2\sqrt{\pi}\Delta{z}^{2}}
+\Omega\sigma\Big),&\quad\frac{\Delta{z}}{\sigma}\gg1,\frac{L}{\sigma}\ll1\;;\\
&0,&\quad\frac{\Delta{z}}{\sigma}\gg\frac{L}{\sigma}\gg1\;;
 \end{aligned} \right.
\end{equation}
In the case of $L/\sigma\ll1$, the concurrence is now  a decreasing function of $\Delta{z}$, which is in contrast to the case of the near zone to the boundary where the concurrence is a growing function of $\Delta{z}$ (see Eq.~(\ref{Expression-c1})). Therefore, one can infer that the amount of harvested entanglement  ought to peak at a certain distance between the detectors and the boundary. This is what has been demonstrated in the numerical evaluations in Ref.~\cite{Zhjl:2021}.
\subsubsection{Large energy gap}
When the detectors' energy gap is much larger than the  Heisenberg energy ($\Omega\gg1/\sigma$), one can easily deduce from Eq.~(\ref{Expression-X}) that  the concurrence  should be vanishingly small. For such identical detectors near the boundary with not too large separation ($\Delta{z}/\sigma\ll1, L/\sigma\ll1$) the concurrence quantifying the harvested entanglement reads approximately,
\begin{align}\label{Expression-c3}
\mathcal{C}_{P}(\rho_{AB})\approx&\frac{\lambda^{2}e^{-\Omega^{2}\sigma^{2}}\Delta{z}^{2}\sigma}{L^{3}\sqrt{\pi}}\;,
\end{align}
As a result, the harvested entanglement degrade to zero as the energy gap increases.
\begin{figure}[!htbp]
\centering  \subfloat[\scriptsize$\Omega_{A}\sigma=0.10$]
{\label{ConvsL11}\includegraphics[width=0.48\linewidth]{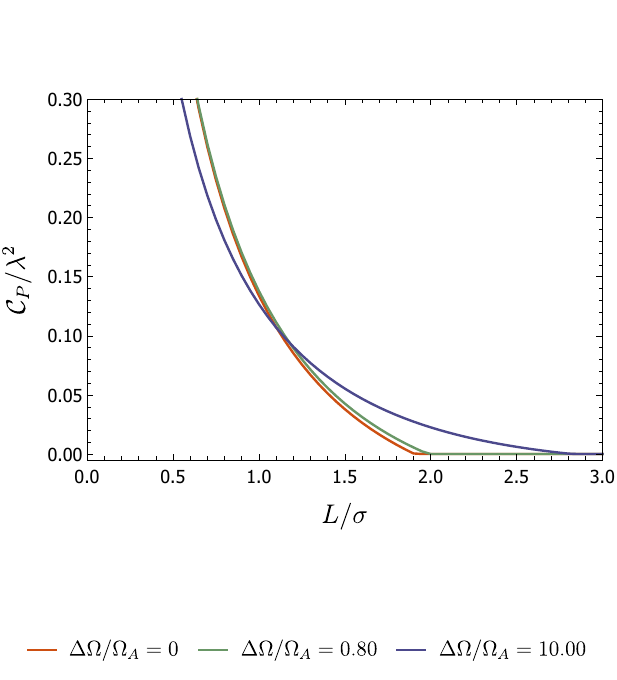}}\;
\subfloat[\scriptsize$\Omega_{A}\sigma=1.10$]
 {\label{ConvsL12}\includegraphics[width=0.48\linewidth]{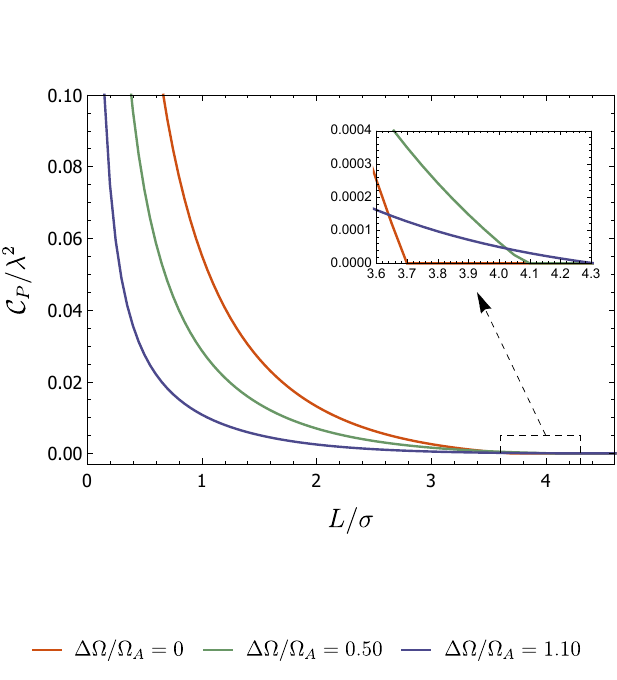}}\;
\caption{The concurrence ${\cal{C}}_P$ as a function of
  interdetector separation $L/\sigma$ is plotted for  $\Omega_{A}\sigma=0.10$, $\Delta\Omega/\Omega_{A}=\{0,0.80,10.00\}$ in (a) and $\Omega_{A}\sigma=1.10$, $\Delta\Omega/\Omega_{A}=\{0,0.50,1.10\}$ in (b) with fixed $\Delta z/\sigma=1.00$.}\label{com-L}
\end{figure}
\subsubsection{Numerical estimation}
\begin{figure}[!htbp]
\centering  \subfloat[\scriptsize$\Omega_{A}\sigma=0.10$]
 {\label{Convsz11}\includegraphics[width=0.475\linewidth]{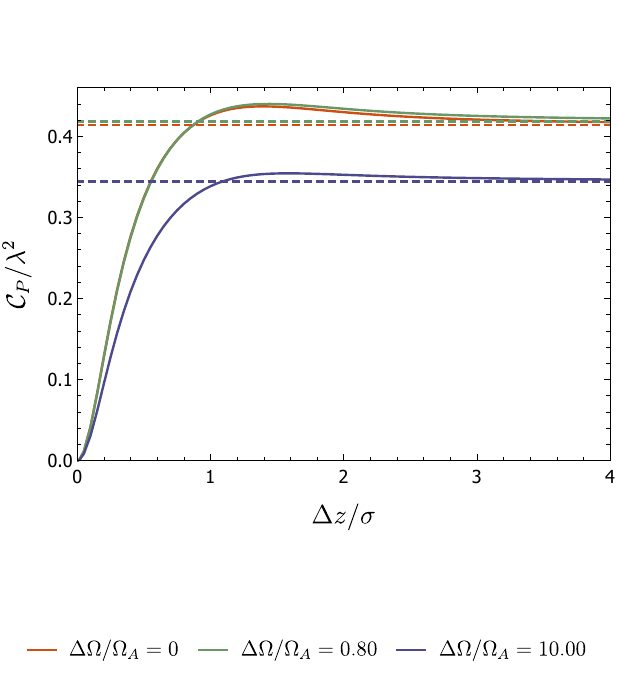}}\;
\subfloat[\scriptsize$\Omega_{A}\sigma=1.10$]
{\label{Convsz12}\includegraphics[width=0.48\linewidth]{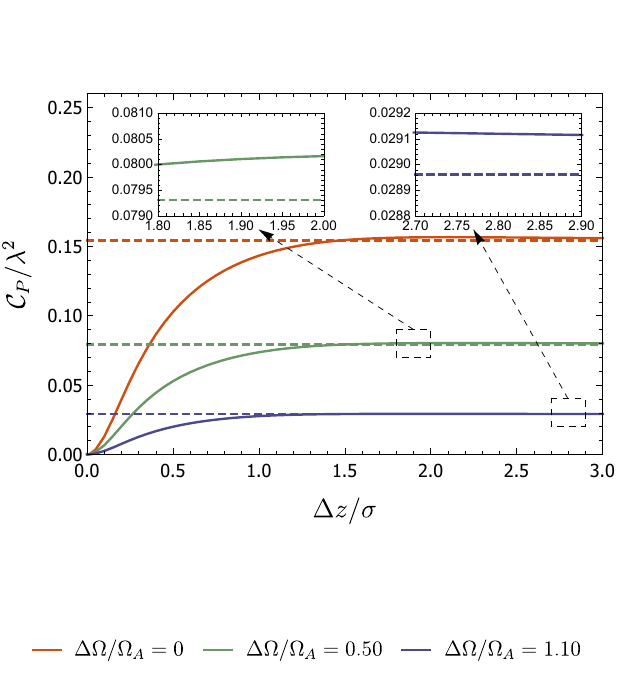}}\\
\caption{The concurrence is plotted as a function of the distance $\Delta z/\sigma$ for fixed $L/\sigma=0.50$. We assume   $\Omega_{A}\sigma=0.10$, $\Delta\Omega/\Omega_{A}=\{0,0.80,10.00\}$ in (a), and $\Omega_{A}\sigma=1.10$, $\Delta\Omega/\Omega_{A}=\{0,0.50,1.10\}$ in (b). All the colored dashed lines correspondingly represent the results in flat  spacetime without any boundaries, i.e.,  the results in the limit of $\Delta{z}\rightarrow\infty$.}\label{com-deltaz1}
\end{figure}

In order to gain a better understanding of  the entanglement harvesting for nonidentical detectors with different energy gaps in the presence of the reflecting boundary, we now  resort to numerical calculation since  analytical approximations are hard to obtain in generic cases. We first show the results from our numerical evaluation on the role played by the reflecting boundary on entanglement harvesting in Figs.~(\ref{com-L}) and~(\ref{com-deltaz1}).

In Fig.~(\ref{com-L}), the  amount of acquired entanglement is  plotted as a function of interdetector separation for various detectors' energy gaps.  Obviously, the harvested entanglement monotonically degrade with the increasing interdetector separation no matter whether the two detectors are identical or not.  For a fixed distance $\Delta{z}$ from the boundary, the identical detectors usually have an advantage in  acquiring more entanglement than nonidentical detectors with unequal energy gaps (e.g., see Fig.~(\ref{ConvsL12})). In contrast, when the energy gaps are small and  the interdetector separation is large enough with respect to the duration time parameter ($\Omega_A\sigma<\Omega_B\sigma\ll1$ and $L>\sigma$), the nonidentical detectors are instead more likely to harvest more entanglement [see  Fig.~(\ref{ConvsL11})].  Notice that there is only a finite  harvesting-achievable range for the interdetector separation for entanglement harvesting,  which can be enlarged by a large  energy gap (or energy gap difference).

 In Fig.~(\ref{com-deltaz1}), we show how the presence of the boundary influences entanglement harvesting by plotting the concurrence  as a function of $\Delta{z}/\sigma$ for various detectors' energy gaps.  It is easy to see that the amount of harvested entanglement is zero in the limit  of $\Delta{z}\rightarrow0$ and approaches to their corresponding values in flat spacetime without any boundaries as $\Delta{z}\rightarrow\infty$ as expected. No matter whether the two detectors are nonidentical with different energy gaps or identical,  the harvested entanglement measure by concurrence  has a peak value when the distance $\Delta{z}$ is comparable to the  duration time $\sigma$ [e.g., see Fig.~(\ref{Convsz11})]. The exact position of detectors where the harvested entanglement peaks is not sensitive to the energy gap difference. Moreover,  the reflecting boundary plays a double-edged role in  entanglement harvesting (inhibiting the entanglement harvesting near the boundary or enhancing it when $\Delta{z}/\sigma>1$ as compared to the case without any boundaries), regardless of whether the detectors are identical or not. Furthermore, both the amount of harvested entanglement and the influence of the boundary can be suppressed by increasing detectors' energy gaps or the energy gap difference between the two detectors.

\begin{figure}[!htbp]
\subfloat[\scriptsize $L/\sigma=0.10$]
 {\label{Convsomega11}\includegraphics[width=0.46\linewidth]{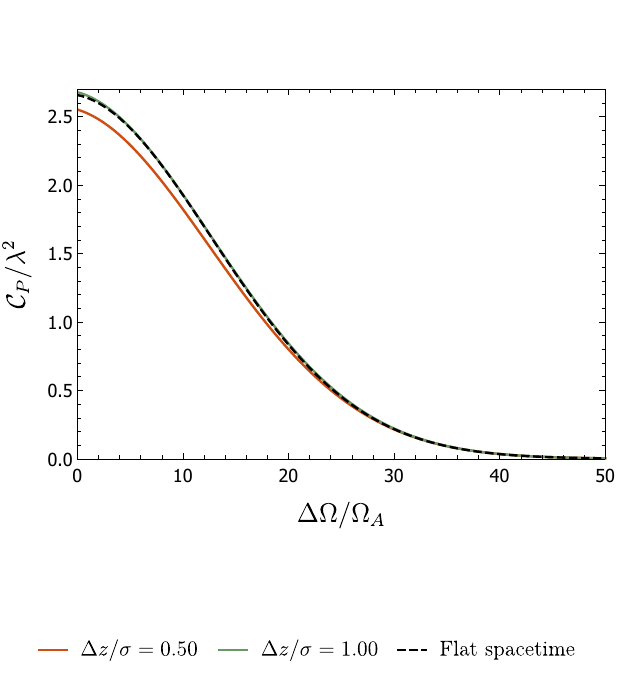}}\;
\subfloat[\scriptsize$L/\sigma=1.50$]
{\label{Convsomega22}\includegraphics[width=0.47\linewidth]{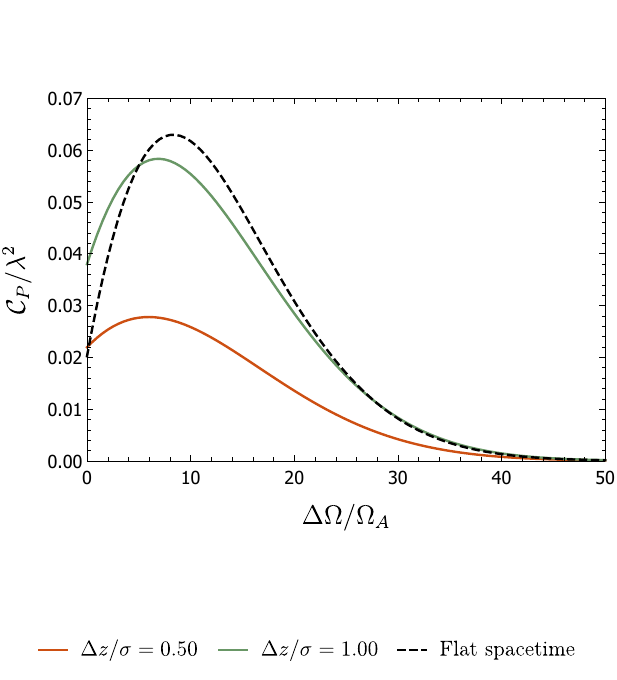}}\;
\caption{The concurrence as a function of $\Delta\Omega/\Omega_{A}$  with $\Omega_{A}\sigma=0.10$ for various $\Delta{z}/\sigma$.  We assume $L/\sigma=0.10$ in (a) and $L/\sigma=1.50$ in (b). Here, the dashed line in all plots indicates the results in flat  spacetime without any boundaries. There is a probability that nonidentical detectors  may harvest more entanglement than the identical ones for large interdetector separation.}\label{com-deltaomega1}
\end{figure}

\begin{figure}[!ht]
  \centering
 \includegraphics[width=0.7\linewidth]{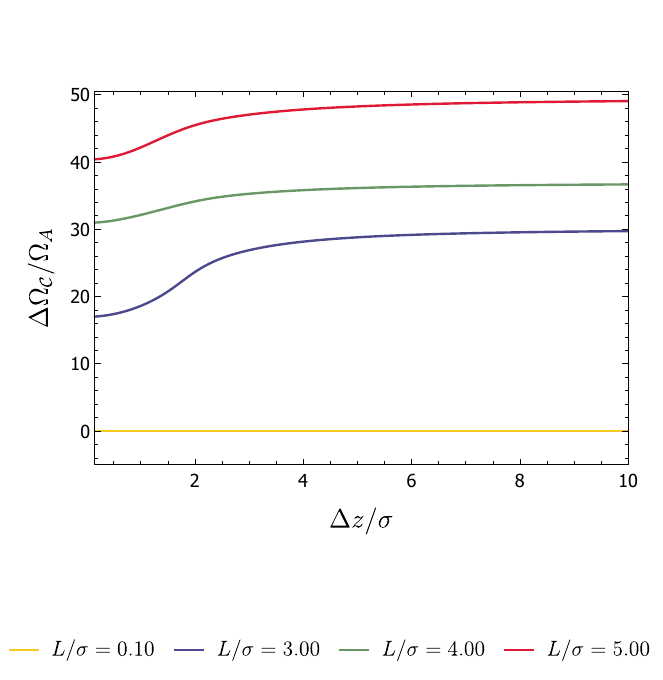}
  \caption{The plot of $\Delta{\Omega_{\mathcal{C}}}/\Omega_{A}$ versus $\Delta{z}/\sigma$ for two detectors aligned parallel to the boundary. Here, we set  $\Omega_A\sigma=0.10$ and $L/\sigma=\{0.10,3.00,4.00,5.00\}$. It is apparent that $\Delta{\Omega_{\mathcal{C}}}/\Omega_{A}$ is an increasing function of $\Delta{z}/\sigma$  as long as the interdetector separation is not too small with respect to the interaction duration time.}\label{DeltaOmegamaxvs-Z2}
 \end{figure}

In order to further study how the entanglement harvesting phenomenon for two nonidentical detectors depends on the energy gap difference, we plot concurrence as a function of the energy gap difference in Fig.~(\ref{com-deltaomega1}). When the interdetector separation $L$ is small relative to the duration time $\sigma$, the energy gap difference would generally hinder  harvesting of entanglement, i.e., the harvested entanglement would rapidly degrade to zero with the increasing energy gap difference between two nonidentical detectors [see Fig.~(\ref{Convsomega11})]. However, when the interdetector separation grows to comparable to or larger than  the interaction duration time parameter, i.e., {$L\gtrsim\sigma$}, the concurrence is no longer a  monotonically decreasing function of the energy gap difference but may initially increase with the increasing energy gap difference, and then reach a maximum value at certain nonzero energy gap difference before degrade to zero as the energy gap difference further increases [see Fig.~(\ref{Convsomega22})].  This suggests that there exist an optimal  energy gap  difference that maximizes the amount of entanglement harvested, which we denote by $\Delta\Omega_{\mathcal{C}}$.  It seems that the value of $\Delta\Omega_{\mathcal{C}}$ depends upon both the interdetector separation and the detector-to-boundary distance. In order to analyze this clearly, we demonstrate how  the value of  the optimal  energy gap  difference is  influenced by the presence of the boundary  for various interdetector separations in Fig.~(\ref{DeltaOmegamaxvs-Z2}). Obviously, the optimal energy gap difference is an increasing function of $\Delta{z}/\sigma$, i.e.,  the presence of the boundary  reduces  the optimal energy gap difference for entanglement harvesting. It is also easy to find the larger the interdetector separation, the larger the optimal value of the energy gap difference, which is consistent with the conclusion in Ref.~\cite{Zhjl:2022.4}.
\subsection{Mutual information harvesting}
Now, we begin to study mutual information harvesting. For identical detectors, the harvested mutual information  can also   be approximated in two cases: small energy gap  and large energy gap.
\subsubsection{Small energy gap}
When the detectors system is placed near the boundary (i.e., $\Delta{z}/\sigma\ll1$), the correlation term $C$ is approximately given by
\begin{equation}\label{Expression-C1}
 C\approx\left\{\begin{aligned}
&\frac{\lambda^{2}\Delta{z}^{2}}{2\pi\sigma^{2}}\Big(\frac{1}{3}-\frac{L^{2}}{15\sigma^{2}}
-\frac{\sqrt{\pi}\Omega\sigma}{2}\Big),&\frac{L}{\sigma}\ll1;\\
&\frac{\lambda^{2}\Delta{z}^{2}}{\pi}\Big(\frac{2\sigma^{2}}{L^{4}}-
\frac{\sqrt{\pi}\Omega}{4\sigma}e^{-\frac{L^{2}}{4\sigma^{2}}}\Big),&\frac{L}{\sigma}\gg1\;,
 \end{aligned} \right.
\end{equation}
then the harvested mutual information can be approximated as
\begin{equation}\label{Expression-I1}
 \mathcal{I}_{P}(\rho_{A B})\approx\left\{\begin{aligned}
&\frac{\ln2}{3}\frac{\lambda^{2}\Delta{z}^{2}}{\pi\sigma^{2}}
+\frac{\lambda^{2}L^{2}\Delta{z}^{2}}{15\pi\sigma^{4}}\ln\Big(\frac{L}{\sigma}\Big)-\frac{\ln2}{2}\frac{\lambda^{2}\Omega\Delta{z}^{2}}{\sqrt{\pi}\sigma}
,&\quad\frac{\Delta{z}}{\sigma}\ll\frac{L}{\sigma}\ll1\;;\\
&\frac{12\lambda^{2}\Delta{z}^{2}\sigma^{6}}{L^{8}\pi}\big(2+3\sqrt{\pi}\Omega\sigma\big),&\quad\frac{\Delta{z}}{\sigma}\ll1,\frac{L}{\sigma}\gg1\;.
 \end{aligned} \right.
\end{equation}
As can be seen from Eq.~(\ref{Expression-I1}), the mutual information harvested by identical detectors, in the limit of $\Delta{z}\rightarrow0$, i.e., the detectors are located at the boundary, must vanish. While, as the detector-to-boundary distance $\Delta{z}$ increases,  the mutual information grows like $\Delta{z}^2$  in the near zone of the boundary. It is also interesting to note that the harvested mutual information  is a decreasing function of the detectors' energy gap $\Omega$ in the case of $L/\sigma\ll1$, while it is an increasing function of the energy gap  in the case of $L/\sigma\gg1$. This is similar to the behavior for the mutual information harvesting in flat  spacetime without any boundaries~\cite{Pozas-Kerstjens:2015}.

For $\Omega\sigma\ll1$ and $\Delta{z}/\sigma\gg1$, the harvested mutual information can be approximated as,
\begin{equation}\label{Expression-I2}
 \mathcal{I}_{P}(\rho_{A B})\approx\left\{\begin{aligned}
&\frac{\lambda^{2}}{2\pi}\Bigg[\ln{2}+\frac{L^{2}}{6\sigma^{2}}\ln\Big(\frac{L}{\sigma}\Big)-\frac{\sigma^{2}}{2\Delta{z}^{2}}\ln{2}
-\sqrt{\pi}\Omega\sigma\ln{2}\Bigg],&\quad\frac{\Delta{z}}{\sigma}\gg1,\frac{L}{\sigma}\ll1\;;\\
&\frac{\lambda^{2}\sigma^{4}}{L^{2}\pi}\Big(\frac{1}{L^{2}}-\frac{1}{2\Delta{z}^{2}}+\frac{\sqrt{\pi}\Omega\sigma}{L^{2}}\Big),&\quad\frac{\Delta{z}}{\sigma}\gg\frac{L}{\sigma}\gg1\;.
 \end{aligned} \right.
\end{equation}
Eq.~(\ref{Expression-I2}) shows that the amount of harvested mutual information is an increasing function of $\Delta{z}$, and its boundary dependence is of  minus $\Delta{z}^{-2}$. The harvested mutual information is also a decreasing function of $\Omega$ in the case of $L/\sigma\ll1$, but becomes an increasing function of $\Omega$ in the case of $L/\sigma\gg1$.
In the limit of $\Delta{z}\rightarrow\infty$,  the Wightman function~(\ref{wigh-2}) reduces to that in flat spacetime, and so the harvested correlations (mutual information or entanglement) approach to the corresponding results in flat spacetime without any boundaries as expected.
\subsubsection{Large energy gap}
When the detectors' energy gap is much larger than the  Heisenberg energy ($\Omega\gg1/\sigma$), the  mutual information for  two identical detectors near the boundary with not too large separation ($\Delta{z}/\sigma\ll1, L/\sigma\ll1$) reads approximately
\begin{align}\label{II33}
 \mathcal{I}_{P}(\rho_{A B})\approx
\frac{\lambda^{2}e^{-\Omega^{2}\sigma^{2}}\Delta{z}^{2}}{8\pi\Omega^{6}\sigma^{10}}\Big[2\Omega^{2}\sigma^{4}\ln2-L^{2}
\ln(\Omega\sigma)\Big]\;.
 \end{align}
One can see that the harvested mutual information degrades to zero as the energy gap increases.
\subsubsection{Numerical estimation}
\begin{figure}[!htbp]
\centering  \subfloat[\scriptsize$\Omega_{A}\sigma=0.10$]
 {\label{MvsL11}\includegraphics[width=0.47\linewidth]{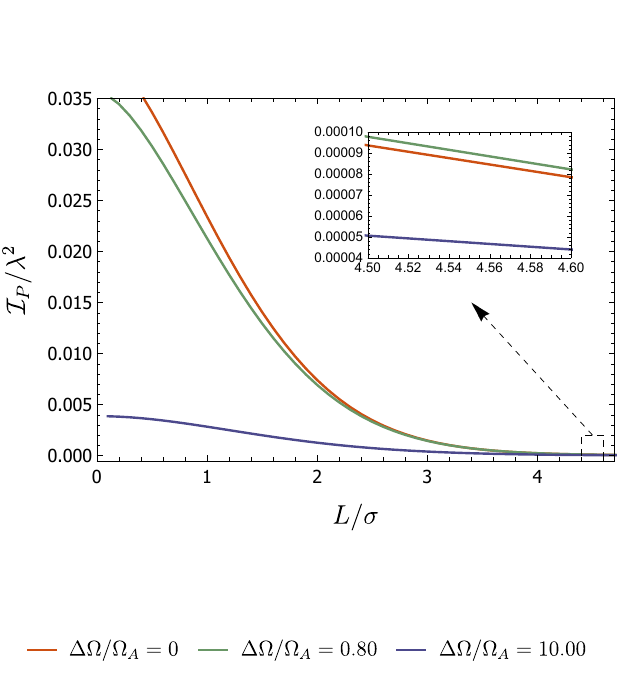}}\;
\subfloat[\scriptsize$\Omega_{A}\sigma=1.10$]
{\label{MvsL12}\includegraphics[width=0.48\linewidth]{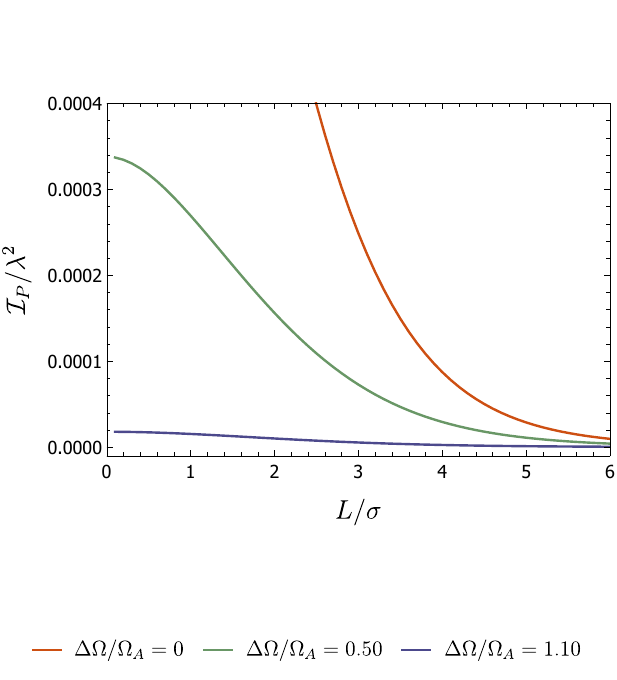}}\;
\caption{ The plots of  mutual information ${\cal{I}}_P $ versus the interdetector separation $L/\sigma$. Here, we have set  $\Omega_{A}\sigma=0.10$, $\Delta\Omega/\Omega_{A}=\{0,0.80,10.00\}$ in (a) and $\Omega_{A}\sigma=1.10$, $\Delta\Omega/\Omega_{A}=\{0,0.50,1.10\}$ in (b) with fixed $\Delta z/\sigma=1.00$.}\label{MvsL}
\end{figure}

\begin{figure}[!htbp]
\centering  \subfloat[\scriptsize$\Omega_{A}\sigma=0.10$]
{\label{Mvsz11}\includegraphics[width=0.48\linewidth]{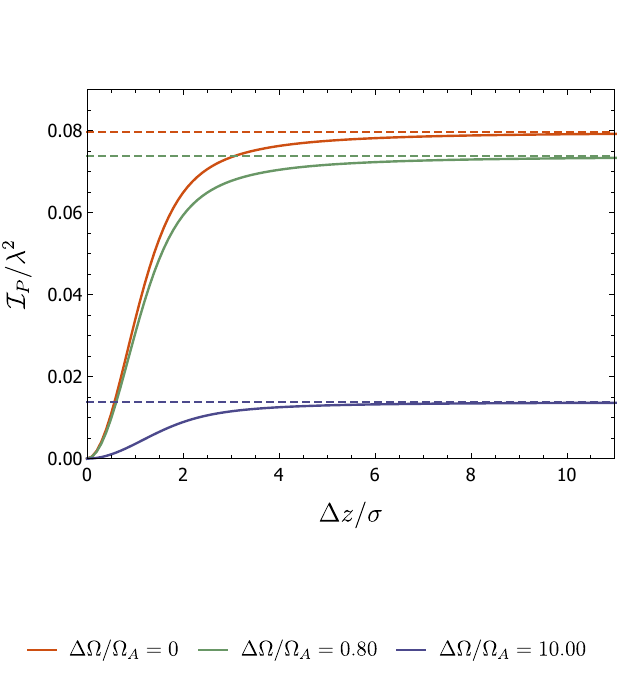}}\;
 \subfloat[\scriptsize$\Omega_{A}\sigma=1.10$]
 {\label{Mvsz12}\includegraphics[width=0.48\linewidth]{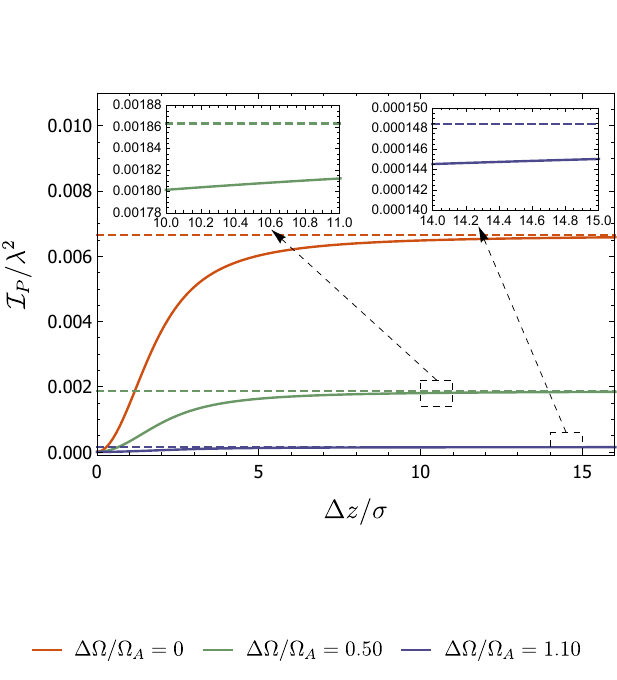}}\;
\caption{The mutual information is plotted as the function of $\Delta z/\sigma$ with $\Omega_{A}\sigma=0.10$, $\Delta\Omega/\Omega_{A}=\{0,0.80,10.00\}$ in (a),  and $\Omega_{A}\sigma=1.10$, $\Delta\Omega/\Omega_{A}=\{0,0.50,1.10\}$ in (b). Here, we have set $L/\sigma=0.50$. The corresponding results in flat spacetime without any boundaries  are shown as dashed lines.}\label{Mvsdeltaz1}
\end{figure}

In more general cases, the mutual information harvesting behavior can be captured by resorting to numerical estimations. In Figs.~(\ref{MvsL}) and~(\ref{Mvsdeltaz1}), the  amount of mutual information is  plotted as a function of interdetector separation and detectors-to-boundary distance for various detectors' energy gaps, respectively. Obviously, similar to the entanglement, the harvested  mutual information is also a monotonically decreasing function of the interdetector separation no matter whether the two detectors are identical or not. Interestingly, when the energy gaps are small and  the interdetector separation is large enough with respect to the duration time parameter ($\Omega_A\sigma<\Omega_B\sigma\ll1$ and $L>\sigma$), the nonidentical detectors with unequal energy gaps are likely to harvest more mutual information than the identical detectors [see Fig.~(\ref{MvsL11})], contrary to  the usual wisdom that the identical detectors may be advantageous in mutual information harvesting. It should be pointed out that, unlike entanglement harvesting, mutual information can be harvested by two static detectors with an arbitrarily large interdetector separation, and physically, this means that the  mutual information harvested outside the harvesting-achievable range of entanglement is either classical correlation or nondistillable entanglement~\cite{Pozas-Kerstjens:2015}.

Fig.~(\ref{Mvsdeltaz1}) demonstrates that in the limit of $\Delta{z}\rightarrow\infty$ the mutual information approaches to its corresponding values in flat spacetime without any boundaries, while on the boundary  the mutual information  vanishes as expected. In sharp contrast to the entanglement harvesting phenomenon which displays a peak-value behaviour when the distance $\Delta{z}$ is comparable to the  duration time $\sigma$ [see Fig.~(\ref{com-deltaz1})], the harvested mutual information is always a monotonically increasing function of $\Delta{z}/\sigma$, approaching asymptotically to the value without any boundaries. So, this means that the reflecting boundary always plays an inhibiting role in mutual information harvesting in contrast to its double-edged role in entanglement harvesting (inhibiting the entanglement harvesting near the boundary and enhancing it when $\Delta{z}/\sigma>1$). Moreover, both the amount of harvested mutual information and the influence of the boundary can be suppressed by increasing detectors' energy gaps or the energy gap difference between the two detectors.

\begin{figure}[!htbp]
\centering  \subfloat[\scriptsize$\Omega_{A}\sigma=0.10$ and $L/\sigma=0.10$]
{\label{Mvsomega11}\includegraphics[width=0.48\linewidth]{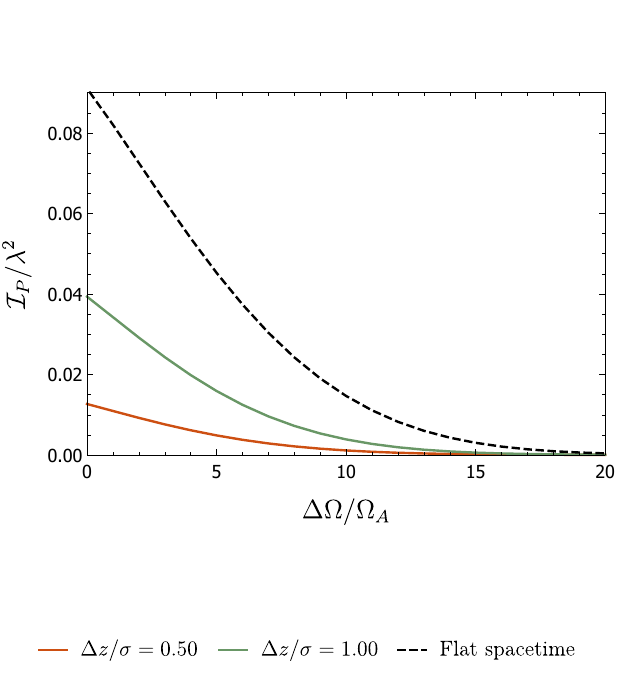}}\;
 \subfloat[\scriptsize$\Omega_{A}\sigma=0.10$ and $L/\sigma=5.00$]
 {\label{Mvsomega22}\includegraphics[width=0.49\linewidth]{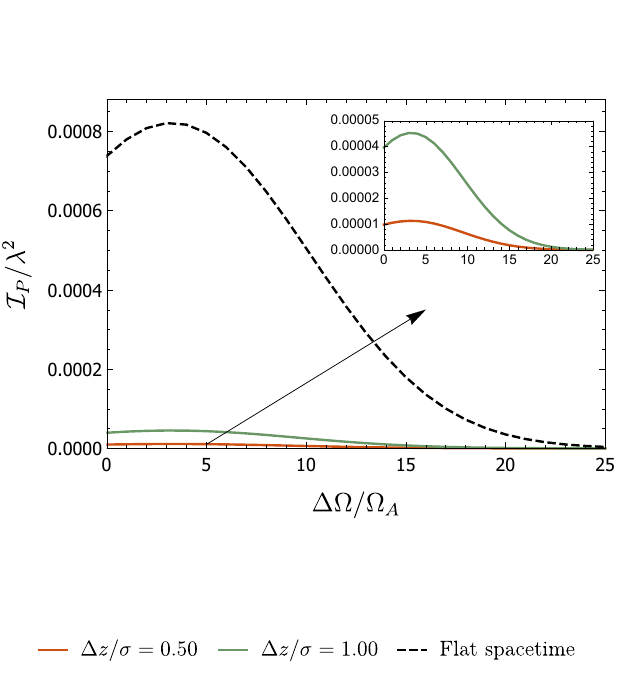}}\;
\caption{Mutual information is plotted as a function of $\Delta\Omega/\Omega_{A}$ for various fixed $\Delta{z}/\sigma$ with $\Omega_{A}\sigma=0.10$ and $L/\sigma=0.10$ in (a) and a spacelike interdetector separation $L/\sigma=5.00$ in (b). Here, the dashed line denotes the  results in flat spacetime without any boundaries. Plot (b) shows that for large interdetector separation the  nonidentical detectors  may harvest more mutual information than identical detectors.}\label{Mvsdeltaomega1}
\end{figure}

\begin{figure}[!ht]
\centering
\includegraphics[width=0.7\linewidth]{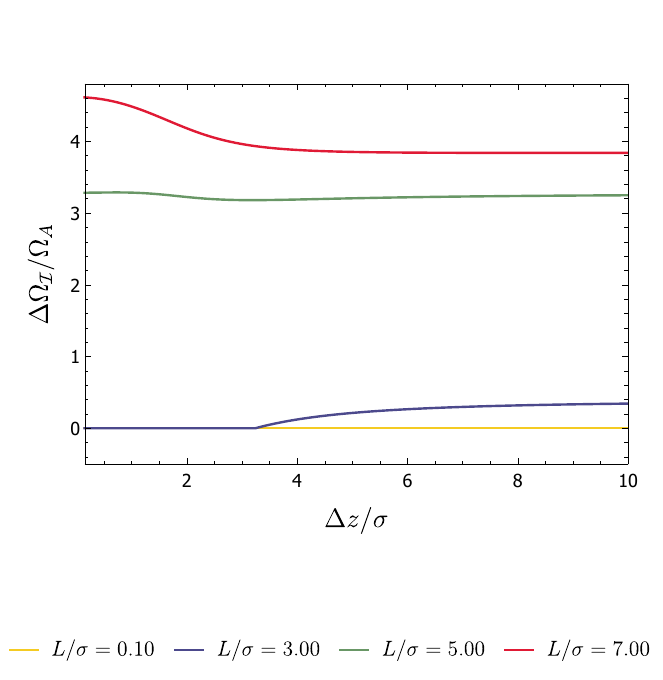}
\caption{The plot of $\Delta{\Omega_{\mathcal{I}}}/\Omega_{A}$ versus $\Delta{z}/\sigma$ with $\Omega_A\sigma=0.10$, $L/\sigma=\{0.10,3.00,5.00,7.00\}$ for two detectors aligned parallel to the boundary. }\label{DeltaOmegamaxvs-Z1}
 \end{figure}

In order to further investigate the influence of the energy gap difference on mutual information harvesting, we plot mutual information as a function of the energy gap difference in Fig.~(\ref{Mvsdeltaomega1}). Similar to entanglement harvesting,  the harvested mutual information is also a monotonically decreasing  function  of the  energy gap difference when the interdetector separation is timelike and very small relative to the duration time $\sigma$ [see Fig.~(\ref{Mvsomega11})]. However, when the interdetector separation is spacelike ($L\gg\sigma$), the mutual information is  no longer a monotonically decreasing functions of the energy gap difference but may  peak  at a certain nonzero energy gap difference and then degrade to zero with the increase of the energy gap difference [see Fig.~(\ref{Mvsomega22})], i.e., there also exists  an optimal energy gap difference that maximizes the mutual information.  We demonstrate how  the  optimal energy gap difference, denoted here by $\Delta{\Omega_{\mathcal{I}}}$, depends upon the presence of the boundary in Fig.~(\ref{DeltaOmegamaxvs-Z1}). Analogous to that in the case of entanglement harvesting, the optimal energy gap difference $\Delta{\Omega_{\mathcal{I}}}$  generally increases as the interdetector separation increases. However, the presence of the boundary can either increase or decrease the optimum energy gap difference for mutual information harvesting.  Specifically, when the interdetector separation is large with respect to the duration time, the optimum energy gap difference for mutual information harvesting is an increasing function of the detector-to-boundary distance; however, when the interdetector separation grows to too large, the optimum energy gap difference for mutual information harvesting becomes a decreasing  function  rather than an increasing  function of the detector-to-boundary distance, which is quite different from  that in entanglement harvesting.

It is well known that once there is a reflecting plane boundary in flat spacetime, the isotropy of spacetime would be lost.  Therefore, the orientation of two detectors system in various angular alignments with respect to the boundary may have a non-negligible impact  on correlation harvesting. So, in the next section,  we will consider the situation in which the two detectors are orthogonally aligned  to the boundary plane.

\section{Correlation Harvesting for the detectors orthogonally aligned  to the boundary}
\begin{figure}[!htbp]
  \centering
  {\includegraphics[width=0.35\linewidth]{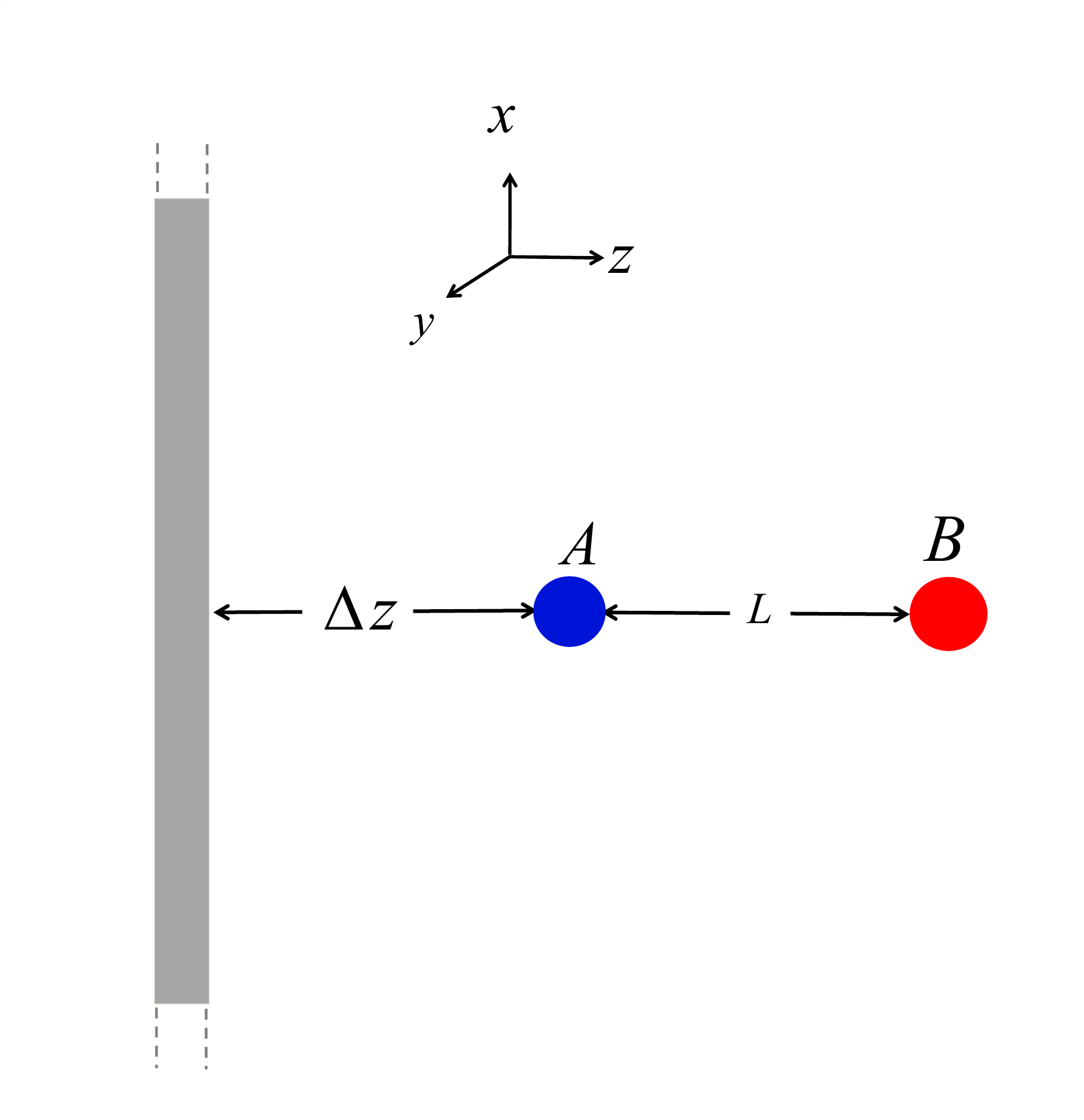}}\qquad
 \caption{ Two static detectors are separated by a distance $L$ orthogonally aligned  to the boundary plane and $\Delta{z}$ is the  distance between the boundary and  the detector which is closer.}
  \label{vermodel}
\end{figure}

The spacetime trajectories of two static detectors orthogonally aligned  to the boundary can be written as [see Fig.~(\ref{vermodel})]
\begin{align}\label{Static-trj2}
&x_A:=\{t_A\;,x=0\;,y=0\;,z=\Delta{z}\}\;,~~&x_B:=\{t_B\;,x=0\;,y=0\;,z=\Delta{z}+L\}\;.
\end{align}
Substituting Eq.~(\ref{wigh-2}) and  Eq.~(\ref{Static-trj2}) into Eq.~(\ref{PAPB}),  we  can see that  $P_A$ is just the expression of Eq.~(\ref{PD0}), and $P_B$ can be obtained by replacing $\Delta{z}$ with $\Delta{z}+L$ in Eq.~(\ref{PD0}). Similarly, the correlation terms $C$ and $X$  can be calculated out
\begin{align}\label{Expression-C11}
    C=&\frac{\lambda^{2}}{4\sqrt{\pi}}
    e^{-\frac{\Delta\Omega^{2}\sigma^{2}}{4}}\Big[f(L)-f(L+2\Delta{z})\Big]\;,
\end{align}
\begin{align}\label{Expression-X11}
 X=-\frac{\lambda^{2}}{4\sqrt{\pi}}e^{-\frac{(2\Omega_A+\Delta\Omega)^{2}\sigma^{2}}{4}}\Big[g(L)-g(L+2\Delta{z})\Big]\;,
\end{align}
with  auxiliary functions $f(L)$ and $g(L)$ defined by Eq.~(\ref{Expression-C0}) and Eq.~(\ref{Expression-X0}), respectively. According to Eq.~(\ref{condf}) and Eq.~(\ref{IIdf}), the  concurrence and harvested mutual information can be directly obtained.
\subsection{Entanglement harvesting}
The approximate expressions for the  concurrence for identical detectors can be also obtained  in some special cases.
\subsubsection{Small energy gap}
For small energy gap ($\Omega\sigma\ll1$), the concurrence, denoted here by $\mathcal{C}_{V}(\rho_{AB})$, can be shown after some long algebraic manipulations to  take the following approximate form
\begin{equation}\label{Expression-cc3}
 \mathcal{C}_{V}(\rho_{AB})\approx\left\{\begin{aligned}
&\frac{\lambda^{2}\Delta{z}}{\sqrt{\pi}}\Big(\frac{\sigma}{L^{2}}+\frac{1}{4\sigma}-
\frac{L}{3\sqrt{\pi}\sigma^{2}}+\frac{L\Omega}{2\sigma}\Big),&\quad\frac{\Delta{z}}{\sigma}\ll\frac{L}{\sigma}\ll1\;;\\
&\frac{\lambda^{2}}{2\sqrt{\pi}}\Big(\frac{\sigma}{L}-\frac{1}{\sqrt{\pi}}+\frac{\sigma^{2}}{2\sqrt{\pi}\Delta{z}^{2}}
+\Omega\sigma\Big),&\quad\frac{\Delta{z}}{\sigma}\gg1,\frac{L}{\sigma}\ll1\;;\\
&0,&\quad{L}\gg{\sigma}\;.
 \end{aligned} \right.
\end{equation}
Similar to the case of parallel-to-boundary alignment, Eq.~(\ref{Expression-cc3}) demonstrates that the entanglement cannot be extracted at extremely large interdetector separation $L$ or vanishing distance $\Delta{z}$ from the boundary.  Comparing Eq.~(\ref{Expression-cc3}) with Eqs.~(\ref{Expression-c1}) and~(\ref{Expression-c2}), we can find for $\Delta{z}/\sigma\ll1$ the harvested entanglement in the orthogonal-to-boundary scenario is also an increasing function of $\Delta{z}$, but the boundary dependence behaves like $\Delta{z}$ rather than $\Delta{z}^{2}$, while for $\Delta{z}/\sigma\gg1$ the approximate form of concurrence in both the orthogonal-to-boundary and parallel-to-boundary alignment looks the same and behaves as a decreasing function of $\Delta{z}$.
This non-monotonicity of a function with respect to $\Delta{z}$ implies that the harvested entanglement in the orthogonal-to-boundary  case also possesses a peak behavior. Beside these, one may find that the detectors orthogonally aligned  to boundary  can harvest comparatively more entanglement than these in parallel-to-boundary alignment for the case of $\Delta{z}\ll L\ll\sigma$. This is physically in accordance with our finding the boundary inhibits the entanglement harvesting for the detectors close to it.
\subsubsection{Large energy gap}
For large energy gap ($\Omega\gg1/\sigma$), the concurrence in the case of $\Delta{z}/\sigma\ll1$ and $L/\sigma\ll1$ can be approximated as
\begin{equation}\label{Expression-c4}
\mathcal{C}_{V}(\rho_{AB})\approx\frac{\lambda^{2}e^{-\Omega^{2}\sigma^{2}}\Delta{z}\sigma}{L^{2}\sqrt{\pi}}\;.
\end{equation}
Similar to the case of the parallel-to-boundary alignment, the harvested entanglement in the orthogonal-to-boundary alignment also degrade with the  increasing detectors' energy gap.

\subsubsection{Numerical estimations}
From above approximations,  it is easy to infer  that the influence of the boundary on the entanglement harvesting phenomenon  for two detectors orthogonally aligned to the boundary should be similar to that for  two detectors aligned parallel to the boundary, i.e., qualitatively, the reflecting boundary should play a double-edge role (inhibiting entanglement  harvesting  for $\Delta{z}\ll\sigma$ and assisting entanglement harvesting  for $\Delta{z}>\sigma$), and moreover, entanglement harvesting  possesses a finite harvesting-achievable range regardless of the presence of the boundary. However, the quantitative details should be slightly different. To better understand it,  we define the difference of concurrence between the scenarios of orthogonal-to-boundary and parallel-to-boundary alignments: $\Delta\mathcal{C}:=\mathcal{C}_{V}-\mathcal{C}_{P}$.

\begin{figure}[!htbp]
\centering \subfloat[\scriptsize$\Omega_{A}\sigma=0.50$ and $L/\sigma=0.50$]
 {\label{Deltacvsz01}\includegraphics[width=0.48\linewidth]{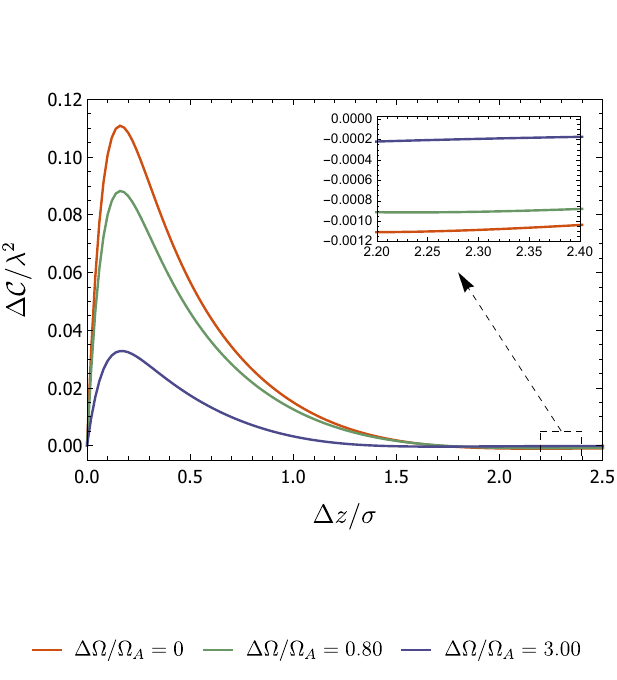}}\;
\subfloat[\scriptsize$\Omega_{A}\sigma=0.50$ and $L/\sigma=3.00$]
{\label{Deltacvsz02}\includegraphics[width=0.48\linewidth]{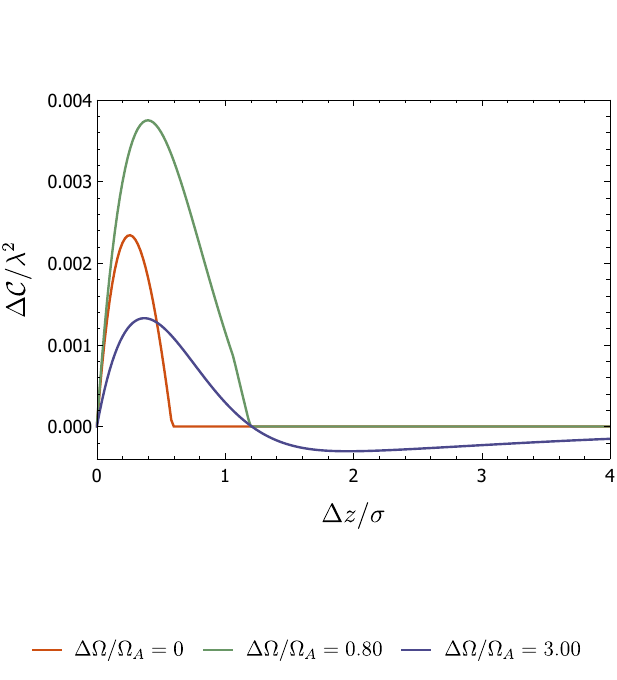}}\\
\caption{The  concurrence difference $\Delta{\cal{C}}$  between orthogonal-to-boundary  and parallel-to-boundary alignments, is plotted as a function of the distance $\Delta z/\sigma$. Here, we have fixed $L/\sigma=0.50$ in plot (a) and $L/\sigma=3.00$ in plot (b). In all plots, we have set $\Omega_{A}\sigma=0.50$ with $\Delta\Omega/\Omega_{A}=\{0,0.80,3.00\}$.}\label{deltac-deltaz2}
\end{figure}

The difference in the amount  of harvested entanglement is plotted as a function of the distance from the boundary in Fig.~(\ref{deltac-deltaz2}).  As we can see from Fig.~(\ref{deltac-deltaz2}), the detectors in orthogonal-to-boundary alignment near the boundary ($\Delta {z}/\sigma\ll1$) could harvest comparatively more entanglement than that in parallel-to-boundary alignment. However, when the detectors are located far from the boundary ($\Delta {z}/\sigma\gg1$) with not too large interdetector separation, this would be reversed,  i.e., the parallel-to-boundary alignment turns out to be favorable for entanglement harvesting. This is because the boundary would play a strongly inhibiting role in entanglement harvesting for two detectors placed near the boundary, and the detectors system in orthogonal-to-boundary alignment  has a relatively longer effective distance from the boundary and thus less  inhibiting effect, resulting in more entanglement harvested  in comparison with those in parallel-to-boundary alignment. While when two detectors are far away from the boundary, the  assisting role played by the boundary leads to more entanglement harvested by detectors in parallel-to-boundary alignment in comparison to orthogonal-to-boundary alignment.

\begin{figure}[!ht]
\centering
\includegraphics[width=0.65\linewidth]{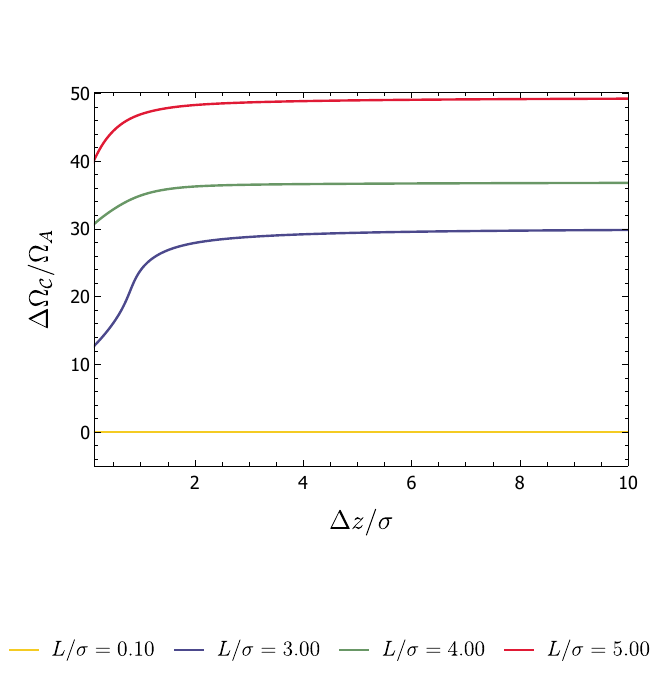}
\caption{The plot of $\Delta{\Omega_{\mathcal{C}}}/\Omega_{A}$ versus $\Delta{z}/\sigma$ with $\Omega_A\sigma=0.10$ and $L/\sigma=\{0.10,3.00,4.00,5.00\}$ for two detectors in orthogonal-to-boundary alignment.}\label{DeltaOmegamaxvs-Z4}
\end{figure}

In Fig~(\ref{DeltaOmegamaxvs-Z4}), we demonstrate how the optimal energy gap difference  that maximizes  the amount of  harvested entanglement depends upon the detector-to-boundary  distance. It is easy to find that the optimal energy gap difference $\Delta{\Omega_{\mathcal{C}}}$ in  orthogonal-to-boundary alignment  is also an increasing function of the detector-to-boundary distance, but the quantitative detail is slightly different from that in the scenario of parallel-to-boundary alignment [comparing  Fig~(\ref{DeltaOmegamaxvs-Z4}) and Fig.~(\ref{DeltaOmegamaxvs-Z2})].

\subsection{Mutual information harvesting}
Now we are in a position of estimating  the mutual information harvested by detectors orthogonally aligned to the boundary. Similarly, the analytical approximation of mutual information can also be obtained for identical detectors  in some special cases.

\subsubsection{Small energy gap}
For small energy gap ($\Omega\sigma\ll1$), the mutual information denoted here by $\mathcal{I}_{V}(\rho_{AB})$ in orthogonal-to-boundary alignment
can be approximated as
\begin{equation}\label{Expression-I3}
 \mathcal{I}_{V}(\rho_{AB})\approx\left\{\begin{aligned}
&\frac{\lambda^{2}\big(2-3\sqrt{\pi}\Omega\sigma\big)\Delta{z}^{2}}{6\pi\sigma^{2}}
\ln\Big(\frac{L}{\Delta{z}}\Big),&\quad\frac{\Delta{z}}{\sigma}\ll\frac{L}{\sigma}\ll1\;;\\
&\frac{32\lambda^{2}\big(1+\sqrt{\pi}\Omega\sigma\big)\sigma^{4}\Delta{z}^{2}}{L^{6}\pi}\ln\Big(\frac{\sigma}{\Delta{z}}\Big),&\quad\frac{\Delta{z}}{\sigma}\ll1,\frac{L}{\sigma}\gg1\;;\\
&\frac{\lambda^{2}}{2\pi}\Bigg[\ln{2}+\frac{L^{2}}{6\sigma^{2}}\ln\Big(\frac{L}{\sigma}\Big)-\frac{\sigma^{2}}{2\Delta{z}^{2}}\ln{2}
-\sqrt{\pi}\Omega\sigma\ln{2}\Bigg],&\quad\frac{\Delta{z}}{\sigma}\gg1,\frac{L}{\sigma}\ll1\;;\\
&\frac{\lambda^{2}\sigma^{4}}{L^{2}\pi}\Big(\frac{1}{L^{2}}-\frac{1}{2\Delta{z}^{2}}+\frac{\sqrt{\pi}\Omega\sigma}{L^{2}}\Big),&\quad\frac{\Delta{z}}{\sigma}\gg\frac{L}{\sigma}\gg1\;.
 \end{aligned} \right.
\end{equation}
In comparison with Eq.~(\ref{Expression-I1}),  the  boundary dependence of mutual information, in the case $\Delta{z}/\sigma\ll1$, is $\Delta{z}^{2}\ln{(\sigma/\Delta{z})}$ in orthogonal-to-boundary alignment, rather than $\Delta{z}^{2}$ in parallel-to-boundary alignment. Hence, the detectors orthogonally  aligned to the boundary  would harvest more mutual information than those aligned parallel to the boundary in the near zone of the boundary. Physically, we can understand it as follows. Since the reflecting boundary always inhibits the harvesting of mutual information as previously discovered,  the orthogonal-to-boundary alignment has a comparatively  longer effective detector-to-boundary distance than the parallel-to-boundary alignment, so  the detectors orthogonally aligned to the boundary could harvest more mutual information.
When compared with Eq.~(\ref{Expression-I2}) for  $\Delta{z}/\sigma\gg1$, the approximated expressions of Eq.~(\ref{Expression-I3}) have  the same form as that of Eq.~(\ref{Expression-I2}), which in the limit of $\Delta{z}\rightarrow\infty$ reduce to the result in flat Minkowski spacetime, which is isotropic, as expected.
\subsubsection{Large energy gap}
For large energy gap ($\Omega\gg1/\sigma$), the mutual information in the case of $\Delta{z}/\sigma\ll1$ and $L/\sigma\ll1$ can be approximated as
\begin{equation}\label{II44}
 \mathcal{I}_{V}(\rho_{AB})\approx \frac{\lambda^{2}e^{-\Omega^{2}\sigma^{2}}\Delta{z}^{2}}{4\pi\Omega^{4}\sigma^{6}}\ln(L/\Delta{z})\;.
\end{equation}
Obviously, the harvested mutual information in orthogonal-to-boundary alignment will degrade to zero as the energy gap grows.
\subsubsection{Numerical estimations}
With the above discussions, it is easy to infer that the influence of the boundary on the mutual information harvesting phenomenon  for two detectors orthogonally aligned  to the boundary should be  similar to that for  two detectors aligned parallel to the boundary, i.e., qualitatively, the reflecting boundary should always play an inhibiting role in mutual information harvesting,  and moreover mutual information can be harvested at  arbitrarily large interdetector separation. However, the quantitative details may in general  be slightly different. To better understand it, we define the difference of mutual information between the scenarios of orthogonal-to-boundary and parallel-to-boundary alignments: $\Delta\mathcal{I}:=\mathcal{I}_{V}-\mathcal{I}_{P}$.

\begin{figure}[!htbp]
\centering  \subfloat[\scriptsize$\Omega_{A}\sigma=0.50$ and $L/\sigma=0.50$]
{\label{DeltaMvsz01}\includegraphics[width=0.48\linewidth]{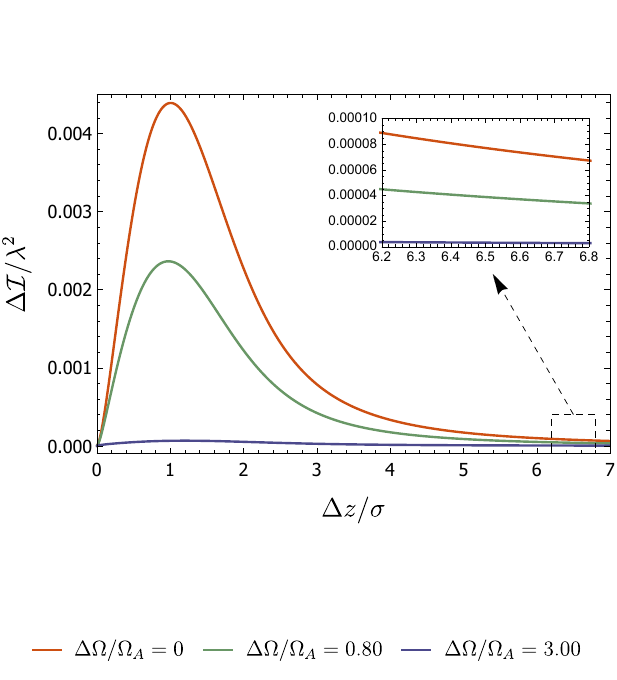}}\;
\subfloat[\scriptsize$\Omega_{A}\sigma=0.50$ and $L/\sigma=3.00$]
 {\label{DeltaMvsz02}\includegraphics[width=0.48\linewidth]{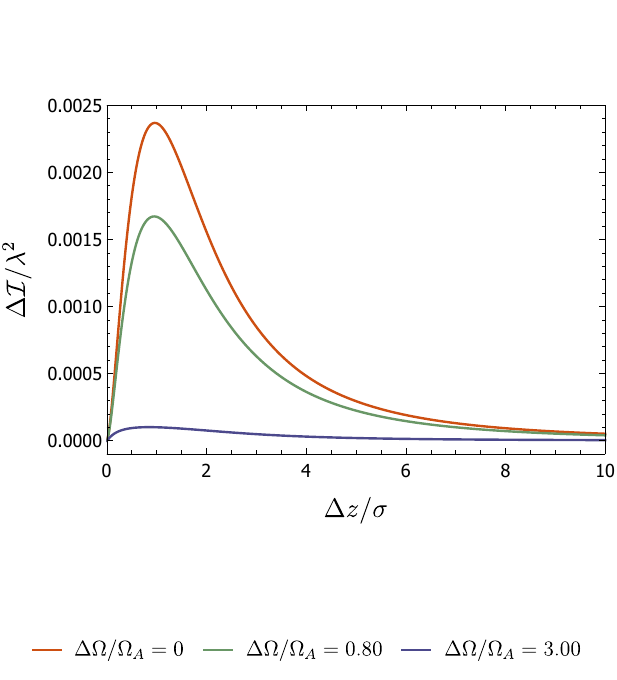}}\;
\caption{The mutual information difference $\Delta{\cal{I}}$  is plotted as a  function of the distance $\Delta z/\sigma$. Here, we have fixed $L/\sigma=0.50$ in plot (a) and $L/\sigma=3.00$ in plot (b). In all plots, we have set $\Omega_{A}\sigma=0.50$ with $\Delta\Omega/\Omega_{A}=\{0,0.80,3.00\}$.}\label{deltam-deltaz2}
\end{figure}

The difference  in the amount of harvested mutual information is plotted, in Fig.~(\ref{deltam-deltaz2}), as a function of the distance between the boundary and the detector which is closer. As we can see from Fig.~(\ref{deltam-deltaz2}),  the case  of orthogonal-to-boundary alignment in general is more favorable for mutual information  harvesting  than that of parallel-to-boundary alignment. This is because the boundary always plays an inhibiting role in mutual information harvesting and the case of orthogonal-to-boundary alignment has relatively longer effective distance from boundary.

\begin{figure}[!ht]
\centering
\includegraphics[width=0.7\linewidth]{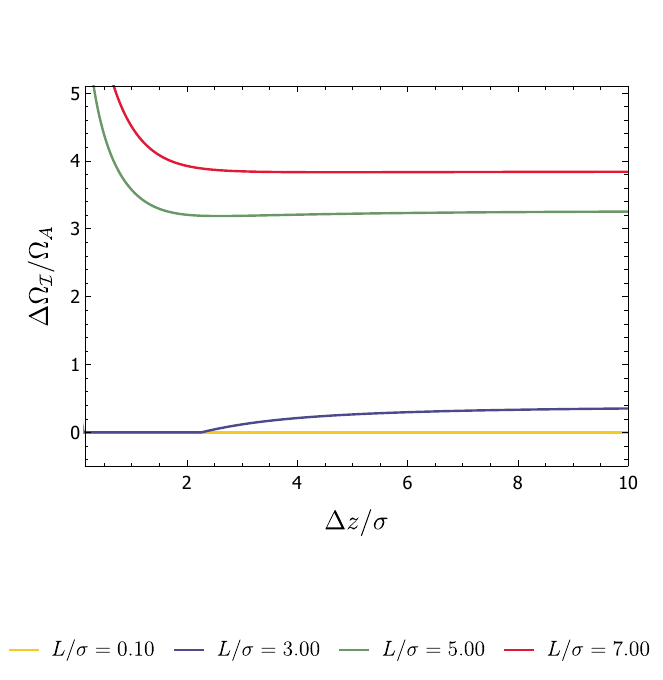}
\caption{The plot of $\Delta{\Omega_{\mathcal{I}}}/\Omega_{A}$ versus $\Delta{z}/\sigma$ with $\Omega_A\sigma=0.10$ and $L/\sigma=\{0.10,3.00,5.00,7.00\}$ for two detectors in orthogonal-to-boundary alignment.}\label{DeltaOmegamaxvs-Z3}
\end{figure}

In addition, the boundary dependence for the optimal energy gap difference between two nonidentical detectors that maximizes the amount of mutual information is displayed in Fig.~(\ref{DeltaOmegamaxvs-Z3}). Similar to the case of parallel-to-boundary alignment,  the optimal energy gap difference $\Delta{\Omega_\mathcal{I}}$ in the case of orthogonal-to-boundary alignment can be either an increasing or a decreasing function of detector-to-boundary distance $\Delta{z}$.  Specifically, when the interdetector separation is large with respect to the duration time, the optimum energy gap difference $\Delta{\Omega_\mathcal{I}}$ is an increasing function of $\Delta{z}$; however, when the interdetector separation is too large, $\Delta{\Omega_\mathcal{I}}$  becomes a decreasing function of $\Delta{z}$.

\section{conclusion}

We  have performed a detailed study on the phenomenon of  correlations harvesting by a pair of UDW detectors interacting with vacuum scalar fields in  flat spacetime with the presence of a perfectly reflecting boundary. The influences of the energy gaps, the energy gap difference and the presence of the boundary  on the mutual information harvesting and entanglement harvesting for nonidentical detectors  are examined  in the scenarios of parallel-to-boundary and orthogonal-to-boundary alignments.

For the case of parallel-to-boundary alignment, both the harvested entanglement and mutual information decrease as the interdetector separation increases, and they are also exponentially suppressed by  increasing  the detectors' energy gaps. It is worth pointing out the entanglement harvesting phenomenon possesses only a finite harvesting-achievable range for interdetector separation, which can  be enlarged by increasing the detectors' energy  gaps (or energy gap difference), while mutual information, in contrast, can be extracted at an arbitrary interdetector separation with an  infinite harvesting-achievable range.
Moreover, when detectors are located near the boundary ($\Delta{z}\rightarrow0$), both mutual information harvesting and entanglement harvesting are inhibited. However, as the distance to the boundary grows to infinity, the harvested mutual information always increases, finally approaching to the corresponding value in flat spacetime without any boundaries. In contrast, the harvested entanglement will surpass the corresponding value in flat spacetime without any boundaries at a certain distance and peak when the distance  becomes comparable to the duration time parameter before turn to decrease and ultimately approach to the corresponding value in flat spacetime without any boundaries when the distance becomes infinite, no matter whether the two detectors are identical or not.
Hence, the reflecting boundary, in  the sense of correlations harvesting, always plays an inhibiting role in mutual information harvesting  but a double-edged role in entanglement harvesting in contrast.

When the interdetector separation is small with respect to the duration time,  the harvested correlations are a monotonically decreasing function of the energy gap difference no matter whether the boundary exists or not, i.e., identical detectors will be advantageous to extracting correlations as compared to nonidentical detectors with different energy gaps. However, for not too small interdetector separation,  the amount of both harvested entanglement and  mutual information  possess a maximum value at a certain nonzero energy gap difference, suggesting that there is an optimal value of energy gap difference between two nonidentical detectors which renders the detectors to harvest most correlations. The optimal energy gap difference generally increases as the interdetector separation increases. Interestingly, the presence of the boundary always decreases the value of the optimal energy gap difference for entanglement harvesting, while it can either increase or decrease the optimum energy gap difference for mutual information harvesting. To be specific, the optimum energy gap difference for mutual information harvesting may be decreased by the presence of the boundary when the interdetector separation is large  with respect to the duration time, and  be increased when the interdetector separation grows to too large.

As for the orthogonal-to-boundary alignment case, the influence of energy gaps and the boundary on both entanglement harvesting and  mutual information  harvesting is qualitatively  similar to that for parallel-to-boundary alignment case, and only  quantitative details are slightly different. Specifically, the detectors  system in orthogonal-to-boundary alignment always harvest comparatively more mutual information than the parallel-to-boundary ones, while they  harvest comparatively more entanglement  only near the boundary. This can be understood physically as a result of that the detectors system  in orthogonal-to-boundary alignment  has a  longer effective distance to the boundary.

Finally,  we have explored in detail  only  two alignments of  the detectors with respect to the boundary, i.e., the parallel-to-boundary and  orthogonal-to-boundary alignments, one may wonder what happens when  the detectors are misaligned, i.e.,  when the line drawn through the detectors is neither parallel nor perpendicular to the boundary. In this case, it is easily seen that the effective distance of the detectors system to the boundary would be shorter than that in  the orthogonal-to-boundary alignment but longer than the distance in the  parallel-to-boundary alignment.  Therefore, one may expect that,  on one hand, qualitatively,  the influence of energy gaps and the boundary on both entanglement harvesting and  mutual information  harvesting will be  similar, and on the other hand, quantitatively,  the detectors in the  misaligned scenario would always harvest comparatively more/less mutual information than the parallel-to-boundary/orthogonal-to-boundary alignment, while they could harvest comparatively more/less entanglement than the parallel-to-boundary/orthogonal-to-boundary alignment only near the boundary.

 \begin{acknowledgments}
 This work was supported in part by the NSFC under Grants  No.12075084 and No.12175062, and Postgraduate Scientific Research Innovation Project of Hunan Province under Grant No.CX20220507.
\end{acknowledgments}
\appendix
\section{The derivations of $C$ and $X$ in parallel-to-boundary alignment}\label{appd1}
\subsection{The correlation term $C$}
 Letting $u=\tau$ and $s=\tau-\tau'$,  we  have, after carrying out the integration with respect to $u$ in Eq.~(\ref{ccdef}),
 \begin{align}\label{ccdef-3}
 C=&\lambda^{2}\sqrt{\pi}\sigma e^{-\sigma^{2}(\Omega_{A}-\Omega_{B})^{2}/4}\int_{-\infty}^{\infty}ds e^{-\frac{s^2+2is\sigma^{2}(\Omega_{A}+\Omega_{B})}{4\sigma^2}}W(s)\;.
 \end{align}
Considering the detectors' trajectories~(\ref{Static-trj}), one can obtain the corresponding  Wightman function
 \begin{align}\label{Wightf-2}
  W(s)=-\frac{1}{4\pi^{2}}\Big[\frac{1}{(s-i\epsilon)^{2}-L^{2}}-\frac{1}{(s-i\epsilon)^{2}-L^{2}-4z^{2}}\Big]\;.
\end{align}
Then the correlation term $C$ takes the form
\begin{equation}
C=C_0-C_z\;,
\end{equation}
with
\begin{align}\label{ccdef-41}
 C_0&=-\frac{\lambda^{2}\sigma}{4\pi^{3/2}} e^{-\Delta\Omega^{2}\sigma^{2}/4}
\int_{-\infty}^{\infty} \frac{e^{-s^{2}/(4\sigma^{2})}}{(s-i\epsilon)^{2}-L^{2}}e^{-i(2\Omega_A+\Delta\Omega)\cdot{s}/2}ds\;,
\end{align}
and
\begin{align}\label{ccdef-42}
 C_z&=-\frac{\lambda^{2}\sigma}{4\pi^{3/2}} e^{-\Delta\Omega^{2}\sigma^{2}/4}
\int_{-\infty}^{\infty} \frac{e^{-s^{2}/(4\sigma^{2})}}{(s-i\epsilon)^{2}-L^{2}-4z^{2}}e^{-i(2\Omega_A+\Delta\Omega)\cdot{s}/2}ds\;,
\end{align}
where $\Delta\Omega:=\Omega_B-\Omega_A$.
Using the Sokhotski formula,
\begin{equation}\label{Sf}
\frac{1}{x\pm{i}\epsilon}={\cal{P}}\frac{1}{x}\mp{i}\pi\delta(x)\;,
\end{equation}
we have
\begin{align}\label{ccdef-5}
 C_0=&-\frac{\lambda^{2}\sigma}{4\pi^{3/2}} e^{-\Delta\Omega^{2}\sigma^{2}/4}{\cal{P}}\int_{-\infty}^{\infty} \frac{e^{-s^{2}/(4\sigma^{2})}}{s^{2}-L^{2}}e^{-i(2\Omega_A+\Delta\Omega)\cdot{s}/2}ds\nonumber\\
 &-\frac{i\lambda^{2}\sigma}{4\pi^{1/2}} e^{-\Delta\Omega^{2}\sigma^{2}/4}
\int_{-\infty}^{\infty} e^{-s^{2}/(4\sigma^{2})}e^{-i(2\Omega_A+\Delta\Omega)\cdot{s}/2}{\rm{sgn}}(s)\delta(s^2-L^2)ds\;
\nonumber\\
=&-\frac{\lambda^{2}\sigma{e}^{-\Delta\Omega^{2}\sigma^{2}/4}}{4\pi^{3/2}}{\cal{P}} \int_{-\infty}^{\infty} \frac{e^{-i(2\Omega_A+\Delta\Omega)\cdot{s}/2}e^{-s^{2}/(4\sigma^{2})}}{s^{2}-L^{2}}ds
-\frac{\lambda^{2}\sigma{e}^{-\frac{\Delta\Omega\sigma^{4}+L^{2}}{4\sigma^{2}}}}{4\sqrt{\pi}L}\sin
 \big[\frac{(2\Omega_A+\Delta\Omega){L}}{2}\big]\;,
\end{align}
where $\rm{sgn}(s)$ represents  the sign function.  The integration in the last line of Eq.~(\ref{ccdef-5}) can be performed straightforwardly by using the convolution of  two functions in Fourier transforms, i.e.,
\begin{align}\label{ccdef-4}
&-\frac{\lambda^{2}\sigma}{4\pi^{3/2}} e^{-\Delta\Omega^{2}\sigma^{2}/4}{\cal{P}}\int_{-\infty}^{\infty} \frac{e^{-s^{2}/(4\sigma^{2})}}{s^{2}-L^{2}}e^{-i(2\Omega_A+\Delta\Omega)\cdot{s}/2}ds
\nonumber\\=&\frac{\lambda^2\sigma}{4\sqrt{\pi}L}e^{-(L^{2}+\Delta\Omega^{2}\sigma^{4})/(4\sigma^{2})}\Im \Big[e^{i(2\Omega_A+\Delta\Omega){L}/2}{\rm{Erf}}
\Big(\frac{iL+2\Omega_A\sigma^2+\Delta\Omega\sigma^2}{2\sigma}\Big)\Big].
\end{align}
Then $C_0$ can be expressed in the form
\begin{equation}\label{Expression-C00}
    C_0=\frac{\lambda^{2}}{4\sqrt{\pi}} e^{-\frac{\Delta\Omega^{2}\sigma^{2}}{4}}f(L)\;
\end{equation}
with
\begin{equation}\label{Expression-C01}
f(L):=\frac{{\sigma} e^{-L^2/(4\sigma^{2})}}{L}\Big\{\Im \Big[e^{i(2\Omega_A+\Delta\Omega){L}/2}{\rm{Erf}}
\Big(\frac{iL+2\Omega_A\sigma^2+\Delta\Omega\sigma^2}{2\sigma}\Big)\Big]
    -\sin\Big[\frac{(2\Omega_A+\Delta\Omega)L}{2}\Big]\Big\}\;.
\end{equation}

Comparing the integration in Eq.~(\ref{ccdef-41})  with that in Eq.~(\ref{ccdef-42}), we easily obtain
the expression of $C_z$ from $C_0$ with $L$ replaced by $\sqrt{L^2+4\Delta{z}^2}$
\begin{equation}\label{Expression-C02}
C_z=\frac{\lambda^{2}}{4\sqrt{\pi}} e^{-\frac{\Delta\Omega^{2}\sigma^{2}}{4}}f(\sqrt{L^2+4\Delta{z}^2})\;.
\end{equation}
Therefore, combining Eqs.~(\ref{Expression-C00}) and
(\ref{Expression-C02}) can verify the validity of Eq.~(\ref{Expression-C}).
\subsection{The correlation term $X$}
Upon considering $W(x_{A}(\tau'),x_{B}(\tau))=W(x_{B}(\tau'),x_{A}(\tau))$, the expression of Eq.~(\ref{xxdef}) can be simplified as
\begin{align}\label{xxdef-20}
 X=&-\lambda^{2}\int_{-\infty}^{\infty}du\chi(u)\chi(u-s)e^{-i(\Omega_{A}+\Omega_{B})u}
 \int_{0}^{\infty}ds\Big[e^{i\Omega_{A}s}W(-s)+e^{i\Omega_{B}s}W(-s)\Big]\nonumber\\
 =&-\lambda^{2}\sqrt{\pi}\sigma e^{-\frac{\sigma^{2}(\Omega_{A}+\Omega_{B})^{2}}{4}}
 \int_{0}^{\infty}ds e^{-\frac{s^2}{4\sigma^2}}\Big[
 e^{is\frac{\Omega_{A}-\Omega_{B}}{2}}W(-s)+ e^{-is\frac{\Omega_{A}-\Omega_{B}}{2}}W(-s)\Big]\;,
 \end{align}
where $u=\tau$ and $s=\tau-\tau'$ are introduced.

Inserting the Wightman function~(\ref{Wightf-2}) into Eq.~(\ref{xxdef-20}), we have
\begin{equation}
X=X_0-X_z\;,
\end{equation}
with
\begin{align}\label{xxdef-41}
 X_0:=\frac{\lambda^{2}\sigma}{2\pi^{3/2}} e^{-\sigma^{2}(2\Omega_A+\Delta\Omega)^{2}/4}
\int_{0}^{\infty}\frac{e^{-s^{2}/(4\sigma^{2})}\cos(s\Delta\Omega/2)}{(-s-i\epsilon)^{2}-L^{2}}ds\;,
\end{align}
and
\begin{align}\label{xxdef-42}
 X_z=&\frac{\lambda^{2}\sigma}{2\pi^{3/2}} e^{-\sigma^{2}(2\Omega_A+\Delta\Omega)^{2}/4}\int_{0}^{\infty}\frac{e^{-s^{2}/(4\sigma^{2})}\cos(s\Delta\Omega/2)}{(-s-i\epsilon)^{2}-L^{2}-4\Delta{z}^{2}}ds\;.
\end{align}
Using the Sokhotski formula~(\ref{Sf}),  we find, after carrying out the fourier transformation,  Eq.~(\ref{xxdef-41}) takes the form~\cite{Zhjl:2022.4}
\begin{align}\label{Expression-Xa11}
     X_0=&-\frac{{\sigma}\lambda^{2}e^{-L^2/(4\sigma^{2})}}{4L\sqrt{\pi}}
    e^{-\frac{(2\Omega_A+\Delta\Omega)^{2}\sigma^{2}}{4}}\Big\{\Im\Big[e^{i\Delta\Omega{L}/2}
    {\rm{Erf}}\Big(\frac{iL+\Delta\Omega\sigma^2}{2\sigma}\Big)\Big]
    +i\cos\Big(\frac{\Delta\Omega{L}}{2}\Big)\Big\}\nonumber\\
    =&-\frac{\lambda^{2}}{4\sqrt{\pi}}
    e^{-\frac{(2\Omega_A+\Delta\Omega)^{2}\sigma^{2}}{4}}g(L)\;
\end{align}
with
\begin{equation}
  g(L):=\frac{{\sigma}e^{-L^2/(4\sigma^{2})}}{L}\Big\{\Im\Big[e^{i\Delta\Omega{L}/2}
    {\rm{Erf}}\Big(\frac{iL+\Delta\Omega\sigma^2}{2\sigma}\Big)\Big]
    +i\cos\Big(\frac{\Delta\Omega{L}}{2}\Big)\Big\}\;.
\end{equation}
Similarly, $X_z$ takes the following form
\begin{equation}\label{Expression-X22}
     X_z=-\frac{\lambda^{2}}{4\sqrt{\pi}}
    e^{-\frac{(2\Omega_A+\Delta\Omega)^{2}\sigma^{2}}{4}}g(\sqrt{L^2+4\Delta{z}^2})\;.
\end{equation}
Combining Eq.~(\ref{Expression-Xa11}) with  Eq.~(\ref{Expression-X22}) gives Eq.~(\ref{Expression-X}).

\def\ACP{AIP Conf. Proc.}
\def\AIHP{Ann. Inst. Henri. Poincar\'e}
\def\AJP{Amer. J. Phys.}
\def\AM{Ann. Math.}
\def\AP{Ann. Phys. (N.Y.)}
\def\APJ{Astrophys. J.}
\def\ASS{Astrophys. Space Sci.}
\def\ATMP{Adv. Theor. Math, Phys.}
\def\CJP{Can. J. Phys.}
\def\CMP{Commun. Math. Phys.}
\def\CPB{Chin. Phys. B}
\def\CPC{Chin. Phys. C}
\def\CPL{Chin. Phys. Lett.}
\def\CQG{Classcal Quantum Gravity}
\def\CTP{Commun. Theor. Phys.}
\def\EASPS{EAS Publ. Ser.}
\def\EPJC{Eur. Phys.  J. C.}
\def\EPL{Europhys. Lett.}
\def\GRG{Gen. Relativ. Gravit.}
\def\IJGMMP{Int. J. Geom. Methods Mod. Phys.}
\def\IJMPA{Int. J. Mod. Phys. A}
\def\IJMPD{Int. J. Mod. Phys. D}
\def\IJTP{Int. J. Theor. Phys.}
\def\JCAP{J. Cosmol. Astropart. Phys.}
\def\JGP{J. Geom. Phys.}
\def\JETP{J. Exp. Theor. Phys.}
\def\JHEP{J. High Energy Phys.}
\def\JMP{J. Math. Phys. (N.Y.)}
\def\JPA{J. Phys. A}
\def\JPCS{J. Phys. Conf. Ser.}
\def\JPSJ{J. Phys. Soc. Jap.}
\def\LMP{Lett. Math. Phys.}
\def\LNC{Lett. Nuovo Cim.}
\def\MPLA{Mod. Phys. Lett. A}
\def\NPB{Nucl. Phys. B}
\def\PCAM{Proc. Symp. Appl. Math.}
\def\PCPS{Proc. Cambridge Philos. Soc.}
\def\PDU{Phys. Dark Univ.}
\def\PLA{Phys. Lett. A}
\def\PLB{Phys. Lett. B}
\def\PR{Phys. Rev.}
\def\PRA{Phys. Rev. A}
\def\PRD{Phys. Rev. D}
\def\PRE{Phys. Rev. E}
\def\PRL{Phys. Rev. Lett.}
\def\PRX{Phys. Rev. X}
\def\PRSLA{Proc. Roy. Soc. Lond. A}
\def\PTP{Prog. Theor. Phys.}
\def\PRp{Phys. Rept.}
\def\RMP{Rev. Mod. Phys.}
\def\SB{Sci. Bull.}
\def\SPP{Springer Proc. Phys.}
\def\SRTU{Sci. Rep. Tohoku Univ.}
\def\ZPC{Zeit. Phys. Chem.}

\end{document}